\documentclass{aa}  
\usepackage{graphicx} 
\usepackage{txfonts}
\usepackage{amsmath}
\usepackage{amssymb}
\usepackage{xcolor}
\usepackage{natbib}
\usepackage{float}

\bibpunct{(}{)}{;}{a}{}{,}
\usepackage{orcidlink}
\usepackage{changes}
\usepackage{comment}
\usepackage{soul}

\newcommand\apgt{\ {\raise-.5ex\hbox{$\buildrel>\over\sim$}}\ }
\newcommand{\aplt}{\ {\raise-.5ex\hbox{$\buildrel<\over\sim$}}\ }
\newcommand{\hes}{$\rm HeS$}
\newcommand{\porb}{\mbox {$\rm P_{orb}$}}

\newcommand{\ms}{$M_\odot$}
\newcommand{\msun}{$M_\odot$}
\newcommand{\zsun}{${\rm Z_\odot}$}
\newcommand{\ls}{\mbox {$L_{\odot}$}}

\newcommand{\myr}{\mbox {~${\rm M_{\odot}~yr^{-1}}$}}
\newcommand{\pyr}{\mbox {{\rm yr$^{-1}$}}}

\newcommand{\te}{\mbox {$T_{\mathrm{eff}}$}}
\newcommand{\logte}{\mbox {$\log(T_{\mathrm{eff}})$}}
\newcommand{\mhe}{\mbox {$\mathrm {M_{He}}$}}

\begin{document}

\title{Elusive hot stripped helium stars in the Galaxy
\\ I. Evolutionary stellar models in the gap between subdwarfs and Wolf-Rayet stars } 

   \author{L. Yungelson \orcidlink{0000-0003-2252-430X}
           \inst{1, 2}
          \and
          A. Kuranov \orcidlink{0000-0003-3821-9852}
          \inst{3} 
          \and
          K. Postnov \orcidlink{0000-0002-1705-617X}
          \inst{3}
          \and
          M. Kuranova
          \inst{4}
          \and\
          L.\,M. Oskinova \orcidlink{0000-0003-0708-4414}
          \inst{1}
          \and
          W.-R. Hamann 
          \inst{1}
}        
   \institute{\inst{1}Institut für Physik und Astronomie, Universität Potsdam, Karl-Liebknecht-Str. 24/25, D-14476 Potsdam, Germany,\\ \email{yungelson@astro.physik.uni-potsdam.de} \\
              \inst{2}Institute of Astronomy, Russian Academy of Sciences, 
              48 Pyatnitskaya str., Moscow 119017, Russia \\
              \inst{3}Sternberg Astronomical Institute, M.V. Lomonosov Moscow State University,
               14 Universitetsky pr.,  Moscow 119992, Russia \\    
               \inst{4}  Faculty of Computational Mathematics and Cybernetics, M.V. Lomonosov Moscow State University, 1-52 Leninskiye Gory, Moscow 119991,  Russia}

   \date{Received August 25, 2023; accepted November 29, 2023) }

% 5 {} token are mandatory
 
  \abstract
  {Stellar evolution theory predicts the existence of He-core remnants of the primary components of intermediate-mass close binaries that lost most of their H/He envelopes due to the mass exchange. They are expected to be observed as (1 -- 7)\,\ms\ hot He-rich stars located in the HRD between sdOB and WR-stars. Several thousands of such stars are expected to  exist in the Galaxy, but none of them have been identified so far. } 
{We aim to provide comprehensive predictions of the numbers and fundamental properties of He-stars and their companions in the Galaxy. This is a necessary first step to guide observations, to enable a comparison between evolutionary models and observed populations, and to determine the feedback of He-stars in the Galaxy. }
  { We expanded the previously considered space of parameters of progenitors of He-stars and  applied a population synthesis based on a grid of models computed by the code MESA.}
   {The estimated number of Galactic binaries hosting (1 -- 7)\,\ms\ He-stars is  $\simeq 20\,000$; it  declines to
$\simeq 3\,000$ for mass
 $\apgt$ 2\,\ms. The decisive factor that defines the number of 
 He-stars is runaway mass loss after Roche lobe overflow by primary components, resulting in formation of common envelopes and merger of components. He-stars are much less numerous than expected, since a fraction of close binaries with $M_{1,0} \aplt$(5 - 7)\,\ms\ produce subdwarfs with masses $\lesssim 1$\,\ms. }
  % conclusions heading (optional), leave it empty if necessary 
     {Overwhelming majority of He-stars reside in binaries with an early-type companions and can be identified neither by the UV excess nor by emission features.
The large periods of a significant fraction of binaries hosting stripped stars ($\apgt$ several hundred days) also hamper their discovery.}
  
\titlerunning{Elusive hot helium stars I.}
\authorrunning{L. Yungelson et al.}
\maketitle 

\section{Introduction}
\label{sec:intro}
  
Mass exchange between components in close binary systems may occur at different evolutionary phases. 
In the so-called “case A,” mass exchange happens if the primary component overflows its Roche lobe (RLOF) in the main sequence.
“Case B” mass exchange occurs if RLOF  happens when the hydrogen shell burning is the main energy source of the star, before helium is ignited in the core.
After pioneering works by \citet{1967ZA.....65..251K,1967ZA.....66...58K,1967AcA....17..355P,1969A&A.....3...83K,1969ASSL...13..253R,
1969ASSL...13..217B,1970A&A.....4..428G,1970A&A.....7..150L,1970AcA....20..213Z,1970Ap&SS...6..497H,1972A&A....19..298G,1974MmSAI..45..893D}
it is accepted 
that at the solar metallicity (Z=\zsun),  case B mass transfer results in the formation of a system 
with a He white dwarf (WD) component, if the zero-age main sequence (ZAMS) donor mass is  $\aplt 2.5$\,\ms, or a hot 
($\logte \apgt 4.4$)
stripped helium star (\hes\ star) component, if the donor's ZAMS mass is higher and the mass loss by its
stellar wind does not prevent RLOF. 
As we show below, HeS stars may also result from case A mass exchange if fast rotation of the close binary components is taken into account.

Stripped helium stars are nondegenerate He-cores of stars that retained a $\lesssim 1$\,\ms\ hydrogen-helium envelope, with the chemical abundance profile formed by several processes: the retreat of an H-burning convective core in the main sequence stage, mixing, further mass loss during RLOF, and stellar wind from the post-RLOF remnant.  

A special interest in the  
\hes\ stars stems from the fact that they are likely among the progenitors of both  hydrogen-poor and hydrogen-rich core-collapse supernovae (SNe) if the mass of their
CO core after He-exhaustion exceeds $\approx 1.4 $\,\ms\
\citep{1986A&A...167...61H},  see also 
\citet{1986ApJ...310L..35U,1994Natur.371..227N,1996ASPC...96..419P,2008ASPC..391..359W,2015ApJ...809..131K,2017ApJ...840...10Y,2020A&A...642A.106D}, as well as progenitors of the electron-capture SNe and SNe~Ia \citep{2022A&A...668A.106C}. 
Furthermore, since HeS stars are hot and young, they might be copious sources of ionizing photons in star-forming galaxies \citep{Dionne2006, 2018A&A...615A..78G, Doughty2021}.

In the Galaxy, $M \aplt$ 2\,\msun\ stars with hydrogen-depleted envelopes are usually identified as sdO/B-type subdwarfs 
\citep{1985ApJS...58..661I,1987ApJ...313..727I,
1987fbs..conf..435T,1990PASP..102..912H,1990SvA....34...57T}, while
stripped stars with $M \apgt 7$\,\msun\ are identified with Wolf-Rayet (WR) 
stars\footnote{See  \citet{2019A&A...625A..57H} and \citet{Sander2019} 
for the latest compilation of mass estimates of Galactic WR stars based on the evolutionary tracks for {\it single rotating stars  with $Z=0.014$} \citep{2012A&A...537A.146E}} \citep{1967AcA....17..355P}. The mass gap between 2\,\ms\ and 7\,\ms\ is thought to be filled by HeS stars.  

In the Galaxy, about 6\,000 sdO/B subdwarfs are detected within 
$\aplt$5\,kpc of the Sun \citep{2020A&A...635A.193G}.
The estimates of the total number of WR stars in the Galaxy range from $1200 \pm 100$ \citep{2015wrs..conf...21C} to about 2\,600 \citep{2017PhDT.......418K}, with   669 objects having already been identified\footnote{Live ``Galactic WR Stars Catalog'' \url{http://pacrowther.staff.shef.ac.uk/WRcat.}}. 
However, currently only a dozen sdO stars with estimated periods and  
mass $\gtrsim$ 1\,\ms\ have been found in binaries     
\citep[see Table 9 in][]{2023AJ....165..203W}.
All of them are companions of Be-stars\footnote{Among the objects listed by \citet{2023AJ....165..203W} as sdO stars, the most massive ($(2.4 \pm 0.5)$\,\ms\ is the subdwarf component of $\gamma$\,Cas-type system $\pi$\,Aqr. However, most recently, \citet{2023PASJ...75..177T} classified this object as a WD and estimated its mass as  $(0.51\pm0.01)$\,\ms. \citet{2023ApJ...942L...6G} suggest that compact objects in
the $\gamma$\,Cas-type subgroup of Be-stars are, actually, not 
sdO-stars, but WDs.\\
  Actually, the cumulative distribution of the sdB masses determined  by combining the spectroscopic analysis
with the fit of the SED and {\it Gaia} parallaxes becomes saturated at $M\approx$0.6\,\ms\ 
\citep{2022A&A...666A.182S}. This may point to a different origin of the ``low''- and ``high''-mass subdwarfs.}. Yet, HR~6819 \citep{2020A&A...641A..43B} and NGC~1850~BH1 \citep{2022MNRAS.511L..24E},
with estimated subdwarf masses $\sim$1\,\ms\, may belong to the same type of stars. There are also several semidetached systems with masses of donors about (0.8 -- 0.9)\,\msun, which may be expected to be progenitors of massive subdwarfs, ``cousins'' of \hes\ stars, for example
DQ~Vel \citep{2013A&A...552A..63B},
V495~Cen \citep{2018MNRAS.476.3039R},
HD~15124 \citep{2022MNRAS.516.3602E},
and V1315~Cas \citep{2023MNRAS.524.5749Z}.
Interestingly, only one ``canonical'' subdwarf companion  ($M=0.426 \pm 0.043$\,\ms) to a Be-star 
($M=3.65 \pm 0.48$\,\ms) has been 
directly observed  \citep[$\kappa$~Dra,][]{2022ApJ...940...86K}.

\citet{2020A&A...639L...6S} disentangled the spectrum of the Galactic star LB-1 \citep{2019Natur.575..618L} and 
suggested that it harbors a $1.5 \pm 0.4$\,\ms\ stripped star with a $(7\pm2)$\,\ms\ Be companion. However,  
\citet{2022A&A...660A..17H} have shown that the modeling of the H$_\alpha$ profile in the spectrum of this binary still does not rule out an alternative hypothesis, according to which LB-1 could host a B-star and a black hole.

The Galactic star $\gamma$\,Columbae  is nitrogen-enriched and has \te=15\,500$\pm$340\,K. Given its estimated  $\log(g)=3.3\pm0.01$, the spectroscopically determined mass is $\sim$4\,\ms\ \citep{2022NatAs...6.1414I}. This led  \citet{2022NatAs...6.1414I} to suggest that $\gamma$\,Col is a remnant of an initially  12\,\ms\ component of a binary that was stripped in a common envelope event and that is currently readjusting its structure to become a hot, compact object. However, no traces of a companion have been found so far, which raises questions about this interpretation.

The helium star in the HD\,45166 system, the famous qWR object, was considered as a prototype of a 4\,\ms\ HeS star and used to anchor the theoretical prescription of the mass-loss rate in the evolutionary models. Recently, it was recognized that the qWR star in HD\,45166 is most likely a strongly magnetic merger product ($M=(2.03\pm0.44)$\,\ms\ ) and that its wind is highly affected by the presence of the magnetic field \citep{2023Sci...381..761S}.

Thus, no \hes\ star with $M\approx(2 - 7)$\ms\ has unequivocally been detected in the Galaxy. 

In nearby low-metallicity galaxies, the number of suspected HeS stars is growing. Recently, \citet{ramachandran2023partially} discovered that a double-line spectroscopic binary, SMCSGS-FS~69, in the Small Magellanic Cloud (SMC, $Z\approx0.2Z_\odot$) harbors an object resembling a HeS star. The object is quite massive, with $M_{\rm HeS} = 2.8^{+1.5}_{-0.8}$\,\ms. Its companion is a 
$M_{\rm Be}=17^{+9}_{-7}$\,\ms\ Be-star. \citet{ramachandran2023partially} found a strongly enhanced N-abundance and moderately enhanced He-abundance in the envelope of a HeS star in the SMCSGS-FS~69, suggesting that the stripping was only partial and that the primary retained a significant fraction of its hydrogen envelope. Such an envelope is consistent with
models of stripping at low metallicities \citep{2022A&A...662A..56K}.  Ramachandran et al. infer that the formation of a helium-rich star in SMCSGS-FS~69 was possible either via case A mass exchange 
or very early case B mass exchange, and point out several further candidate \hes\ stars in the Large Magellanic Cloud (LMC, $Z\approx 0.5Z_\odot$.)

\citet{2023arXiv230700061D} measured UV-magnitudes  of $\simeq$500\,000 stars in the direction of the LMC and SMC galaxies and selected 25 stars that may be binaries harboring \hes\ stars. Ten of these were spectroscopically analysed by \citet{2023arXiv230700074G} who determined their effective temperatures,   
$\te\sim$(50\,000 -- 100\,000)\,K, and bolometric luminosities,  
$L/\ls \sim(10^3 - 10^5$.) 
%The stars have  surface gravities 
%$\log(g)\sim$5, and hydrogen-deficient ($X_{\rm H} \simeq$ 0.0 -- 0.4) atmospheres.
Surface gravities of the stars are $\log(g) \approx $5, while hydrogen abundance 
in their atmospheres (by mass)  does not exceed 0.4. 
These characteristics are compatible with those expected for (1 -- 8)\,\ms\ \hes\ stars. 

Furthermore, as a possible explanation for the composite spectrum of the binary VFTS~291 in the 30~Dor complex in the LMC,  \citet{2023MNRAS.525.5121V} suggested that the binary contains a $(2.2\pm0.4)$\,\ms\ HeS star, which is bloated due to instabilities in the He-burning shell. If true, this object must be quite unique, since stars spend less than 1\%\ of their total nuclear-burning lifetime in this stage. 

A systematic study of theoretical HeS star populations aimed at determining spectral and photometric characteristics along a sequence of increasing masses was performed by \citet{2018A&A...615A..78G}. This study renewed the general interest in the topic; however, it was limited to the products of a rather early case B mass exchange. In this work, only one 
model HeS star with a core He abundance of 0.5 was selected from each  evolutionary track with a given initial combination of ZAMS masses of components. The initial ratio of the  primary and secondary masses ($M_1, M_2$) was fixed to $q_0=M_2/M_1$=0.8.  Furthermore, a single value of the initial orbital period was considered. \citet{2018A&A...615A..78G} highlight the importance of recipes prescribing the mass-loss rate by radiatively driven stellar winds for the evolution and spectral appearance of the HeS stars. It should be noted however that their calculations were anchored to the empirically derived mass-loss rate of the qWR component in HD\,45166, which is by now known to be spurious \citep{2023Sci...381..761S}.

In the present study, we computed an extended grid of binary stellar evolutionary models, leading to the formation of objects that could be identified with \hes\ stars, if observed.
For the first time, we studied the entire range of combinations of the primary and secondary masses ($M_1, M_2$) and orbital period
($\porb$) on ZAMS that could produce binaries containing \hes\ stars. We determined the fundamental stellar parameters and the surface helium to hydrogen abundance ratios for 
the \hes\ stars, as well as for their companions. This allowed us to accomplish a population synthesis for  binaries harboring (1 -- 7)\,\ms\ \hes\ stars in the Galaxy, and to evaluate their number and distributions over different parameters. In a subsequent paper, we plan to use the Potsdam Wolf-Rayet (PoWR)
non-LTE code \citep[e.g.,][]{2019A&A...621A..85H} to produce synthetic spectra of
binaries containing HeS stars, with the goal of enabling informative searches of binaries with HeS stars and/or explaining the selection effects precluding the detection of these systems. 

The paper is organized as follows. In Sec.~\ref{sec:model}, the model and its assumptions are introduced. Section~\ref{sec:results} describes the results of the model's calculations, while the discussion of the obtained results and our conclusions are presented in Sec.~\ref{sec:disc}. In the appendix we display the test results, assuming alternative mass-loss prescriptions.
\begin{figure*} %1
\centering
\includegraphics[width=8.4cm]{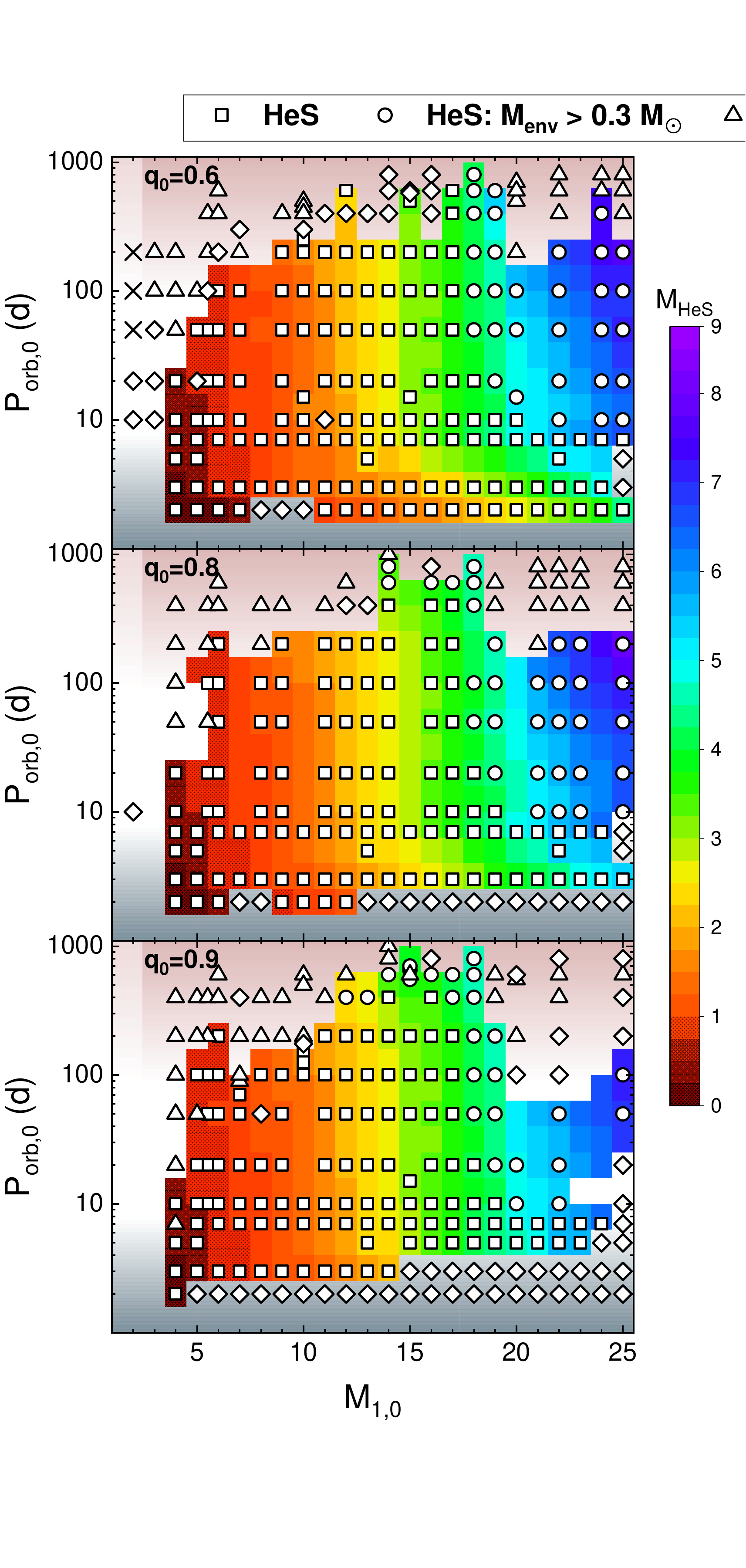}
\hskip -0.089 cm
\includegraphics[width=8.4cm]{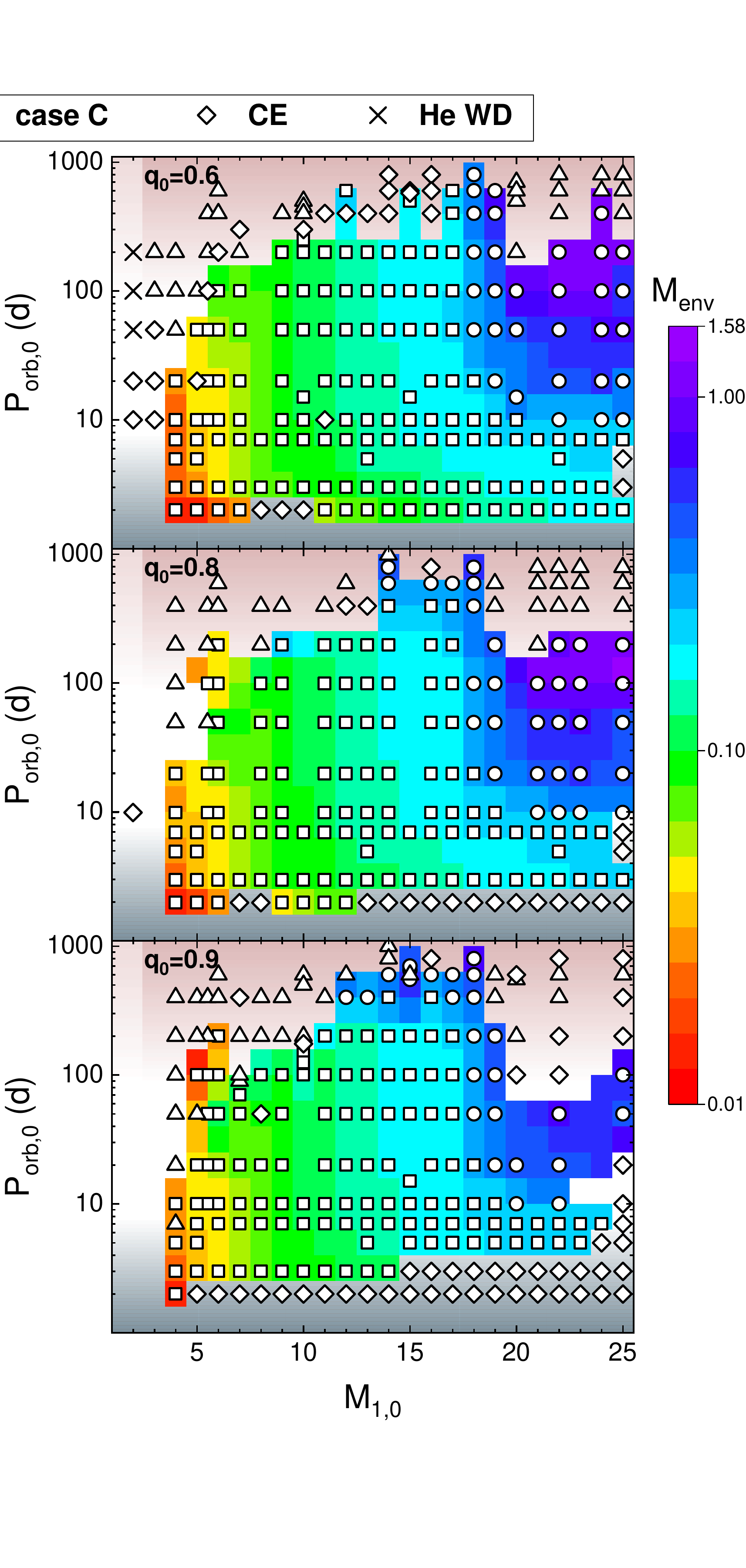}
\caption{Grid of computed systems. {\it Left panel}: Models of binaries 
with initial mass ratios, $q_0=M_{2,0}/M_{1,0}=0.6, 0,8, 0.9$ in the ``ZAMS mass of the primary component, $M_{1,0}$  -- initial 
orbital period, $\porb_{,0}$'' plane. Color-coded are the masses of the remnants of primary components of the binaries after RLOF at the point of the lowest luminosity along the post-mass-exchange evolutionary track, 
obtained by a numerical approximation of the results of the computation of a non-regular grid of models. The legend over the upper panel shows the evolutionary ``fate'' of the computed binaries. Open squares and circles indicate the formation of \hes\ stars (in the latter case, with retained H/He envelopes exceeding 0.3\,\ms.)
The diamonds show binaries entering common envelopes.  Open triangles stand for binaries where case C mass transfer occurred. In the light gray region at the top left of the middle and lower panels, the models for $q_0$=0.9, 0.8 were not computed, since  binaries in this region definitely experience case C mass transfer and form common envelopes, or the \te\ of the stripped component never exceeds 25\,000\,K. Crosses mark the systems forming He WDs. 
The ``pixels'' overplotted by dots   
show the locations of the progenitors of stripped helium remnants with masses $<1$\,\ms.
{\it Right panel}: Color-coded  masses of H/He envelopes retained by the remnants of the initial primary components shown in the left panel at $L_{\rm min}$\ along their tracks (with steps 0.01\,\ms\ and 0.1 \ms\ below and above 
${\rm M_{env}}=0.1$\,\ms\, respectively.
All masses are in \ms.} 
\label{f:M_Porb_full_grid}
\end{figure*} 
\section{The model}
\label{sec:model}

\subsection{Hot stripped helium stars}
Our goal was to find HeS stars with masses bridging the gap between sdOB and WR stars. To achieve this, we computed a grid of models of interacting binaries with the primary masses on ZAMS in the range 
$M_{1,0}  \in [3 - 28]$\,\ms, the initial mass ratios of the components, $q_0=M_{2,0}/M_{1.0}$=0.6, 0.8, 0.9, and initial orbital periods between two days to several hundred days (depending on the masses of components and the initial $q_0$.) In the systems with $q_0 < 0.6$\, the components were expected to merge at RLOF stages; see below. However, we made several test runs for binaries with  $q_0=0.4$ and found that in some cases the merger may be avoided. 

As the first step, we defined the range of stellar parameters of binaries that allows us to identify stellar remnants after RLOF as HeS stars.
We considered two values of the lower mass limit of \hes\ stars -- 1\,\ms\ and 2\,\ms --  in order to also address the scarcity of the observed ``heavy''  subdwarfs. The upper mass limit of  \hes\ stars was taken as 7\,\msun, following \citet{2000A&A...360..227N}, which is  similar to the lowest mass estimate
of the Galactic WR stars belonging to the nitrogen spectral subsequence (WN) obtained by \citet{2019A&A...625A..57H}. 
In our models, the luminosity of a stripped  7\,\ms\ He star in the He shell burning stage corresponds to that of a star with a ZAMS progenitor   
mass close to 24\,\ms\ and agrees also with the lower limit of the luminosities of Galactic, spectroscopically identified WN stars $\log(L/\ls)\approx4.9$
\citep[see Fig.~3 in][]{2020A&A...634A..79S}. This sets an upper mass limit of primary components of the models of close binaries in our computations. However, we also made several runs for binaries with masses up to 28\,\ms.
  
The considered temperature range was  limited to  the ``hot'' objects, with $\log(\te) \geq 4.4$, similar to \te\ of sdOB subdwarfs. It should be noted that 
the stars occupying the uppermost part of the main sequence are hotter (see Figs.~\ref{f:HR_1}, \ref{f:HR_2}  in the Sec.~\ref{sec:results}.)  

In this paper, we considered only one channel for the HeS star formation -- a stable, nonconservative mass transfer in close binaries. This meant that
we omitted other possibilities for the formation of HeS stars, such as a merger during a common envelope or the survival of the progenitor of a 
\hes\ star in common envelopes. All these scenarios have too many additional free parameters while hardly increasing the number of Galactic \hes\ stars. 

\subsection{Model assumptions}
For the computations of the evolutionary tracks, we used the code MESA  \citep{2011ApJS..192....3P,2013ApJS..208....4P,2015ApJS..220...15P,2018ApJS..234...34P,2019ApJS..243...10P}, release 12778.
Computations were performed for the  metallicity Z=0.02.
We applied physical assumptions in the code that are similar to those used by \citet{2020ApJ...903...70S} in their study of SN~IIb progenitors.
In the case of close binaries experiencing case A mass exchange, we accounted for rotation-induced mixing following   
\citet{2022A&A...659A..98S}.

Unlike \citet{2020ApJ...903...70S}, we did not
assume fixed values for the accretion efficiency but instead treated the mass and angular momentum loss from the system as regulated by the critical rotation of the accretor \citep{1981A&A...102...17P}. 
We assumed that the mass transfer through the vicinity of the $L_1$ point is conservative up to the instant
when the initially nonrotating accretor attains a critical equatorial rotational velocity. After that, the accretion rate is limited by the amount of matter that corresponds to the critical rotation of the accretor, while the excess of the accreting matter leaves the system taking away the accretor's specific 
angular momentum. This is a kind of  a ``standard'' scenario of the formation of \hes\ stars in binaries elaborated also by  N. Langer and his coauthors \citep[e.g.,][]{2003IAUS..212..275L,2005A&A...435.1013P,2022A&A...659A..98S} and  applied, 
for instance, by \citet{2010ApJ...725..940Y,2017ApJ...840...10Y} in studies of the progenitors of core-collapse SNe and by \citet{2017A&A...608A..11G} in computations of models of stripped stars. 
The resulting evolution is completely conservative in the sense of mass
and angular momentum before the rotation of the accretor becomes critical, but becomes almost nonconservative in mass and angular momentum later.
\footnote{Efficiency of accretion is uncertain, but, definitely, small.  In \citet{2017ApJ...840...10Y} the authors set efficiency of accretion to 20\%. This causes slight difference of the masses of the RLOF remnants compared to the ones obtained in the present paper, where efficiency varies around 5\% -- 7\%.}.

The main-sequence companions of the nascent \hes\ stars become rapidly rotating and may be identified with Be-stars \citep{1975BAICz..26...65K,1991A&A...241..419P}; see also  
\citet{2022MNRAS.516.3602E} and references therein. This mechanism is consistent with the apparent deficiency of main-sequence companions to Be-stars due to a large difference in visual magnitudes and the difficulty of discovering subdwarf companions of Be-stars in UV \citep{2020A&A...641A..43B}. As El-Badry et al. claim, 10 to 60 percent
of all Be-stars may be formed via this mechanism. 

Other most critical assumptions concern the treatment of 
stellar winds.  
For $\te\leq$10\,000\,K and the surface hydrogen abundance, $X_s \geq 0.4$, we followed the prescription incorporated in MESA 
\citet{1988A&AS...72..259D} for the mass-loss rates over the HR-diagram scaled by (Z/\zsun )=0.85 to match the Z-scaling of \citet{2001A&A...369..574V}. 
The wind mass-loss recipe from the latter paper was used for $\te\geq$11\,000\,K.
For 10\,000\,K $\le \te \le$11\,000\,K, the mass-loss rates were obtained by interpolation.
For $\te \geq 11\,000$\,K and $X_s \leq 0.4$ we used
the mass-loss rates from \citet{2000A&A...360..227N}.

The formation of HeS stars in case A mass exchange deserves special consideration.
The relative number of stars subject to case A mass transfer is small, but not negligible, because the initial distribution of binaries over the logarithm of orbital periods is taken flat. Donors of the most tight binaries can experience mass loss on the main sequence. The components of these binaries are rapid rotators due to tidal effects synchronizing the orbital and axial rotation.
The velocities of the axial rotation in the systems with ${\rm P_{\rm orb, 0}}$\ equal to a few days may amount to several
100\,${\rm km\,s^{-1}}$. The rapid rotation induces a number of instabilities that result in the redistribution of the angular momentum and chemical species inside the stars 
\citep[see][]{2000ApJ...528..368H,2000ApJ...544.1016H}. When accretion starts, the rotational velocity increases because of the angular momentum carried by the accreted matter. 
In order to treat more accurately case A mass exchange, we computed the evolution for this
case following  \citet{2022A&A...659A..98S}, who took the effects of mixing into account\footnote{Note that in the latter paper the diffusion
coefficients describing the rotation-induced instabilities have the maximum values.}.
Some systems experiencing case A mass exchange then avoid merging in the main-sequence and may produce HeS stars.
Since the most rapidly rotating stars are components of low-mass tight binaries, the main effect of the case A evolutionary path is an increase in the number of low-mass He stars with masses overlapping with the mass range of 
``canonical'' sdB/O stars, below 1\,\ms\ and extending to (4 -- 5)\,\ms.  

Close binaries that experience case B  mass exchange avoid merging, unless the initial binary mass ratio is $q_0 \lesssim$ 0.6.

\subsection{Population synthesis and cut factors}
In the population synthesis, we assumed a Galactic star formation rate, SFR=2\,\myr\ \citep{2011AJ....142..197C,2015ApJ...806...96L},
the Salpeter IMF for the primaries of the initial systems,
$dN/dM \propto M^{-2.35}$ between 0.1\,\ms\ and 100\,\ms, a flat distribution over mass ratios of components on ZAMS, $q_0=M_{2,0}/M_{1,0}$, and a
flat distribution over 
$\log(\mathrm{P_{orb,0}})$ \citep{1924PTarO..25f...1O,1982Ap&SS..88...55P}. 
The binarity rate was taken to be 50\% (i.e., 2/3 of stars are in binaries.)
Under these assumptions, the 
number of binaries born annually in the Galaxy is $B\approx 1.14$\,SFR/\ms.

A detailed study of the (3 -- 28)\,\ms\ range of masses of primaries in close binaries with different mass ratios of components
and orbital periods at ZAMS shows that not all stars in this mass range, in which
the H shell burning layer is the main source of the energy release, in other words subject to case B mass exchange, really contribute to the formation of \hes\ stars via the RLOF.
There are several ``guillotine'' factors. 

(i) If the initial system is tight enough, the rejuvenation of the accretor, by bringing accreted matter into the core of 
the star \citep{1977A&A....54..539K}, which leads to an increase in the stellar radius, results in contact between the components and, most probably, the eventual formation of a rapidly rotating single star \citep[e.g.,][]{2011IAUS..272..531D,2021MNRAS.507.5013M}. 
In case A mass exchange, this may happen even in the stage when the former donor contracts to high temperatures, but the former accretor is still a main-sequence star.

(ii) If the star fills its Roche lobe while it has a deep convective envelope, the mass loss proceeds in the dynamical
timescale and leads to the formation of a common envelope, which may result in the merger of the components
or the formation of a tight binary system. 
In addition, if $q\aplt(0.4 - 0.6)$, even if the envelope is radiative, the mass loss typically occurs on a dynamic timescale and leads to the formation of a common envelope.
Modeling of common envelopes requires 3D computations. All attempts to compute the evolution of CE-systems have thus far been unsuccessful, since a lot of processes occurring on different timescales are involved 
\citep[see, e.g., ][]{2016ApJ...816L...9O,2023A&A...674A.121G}. 
Thus, there is an upper limit for the range of ZAMS periods of the potential  precursors of \hes\ stars of several 100\,days and a lower  limit for $q_0$ (see Fig.~\ref{f:M_Porb_full_grid}.)

In conventional population synthesis, the outcome of evolution in common envelopes is treated using the so-called ``common envelope efficiency'' and the binding energy of the donor envelope 
\citep{1984ApJ...277..355W,1990ApJ...358..189D}. These parameters are highly uncertain
\citep[e.g.,][]{2013A&ARv..21...59I}. We discarded systems that pass through the common envelope stage and, therefore, they were 
``lost'' in our modeling as binaries with \hes\ stars. We crudely estimated that the fraction
of the ``lost''  binaries may comprise $\simeq$10\% of the total population of \hes\ stars. 

(iii)  If the potential donor in a close binary is massive enough to ignite He in nondegenerate conditions,
it may happen that, when it fills the Roche lobe between the terminal age main sequence (TAMS) and the base of the red giant branch in HRD, the mass of its 
He-core (in fact, only slightly less massive than the future \hes\ star)  still does not 
exceed $\approx$(1 -- 2)\,\ms\ (Fig.~\ref{f:M_Porb_full_grid}.)  Observationally,  the remnant will probably be identified as a subdwarf. 
This sets a lower limit on the HeS stars' progenitor masses
(5 -- 7)\,\msun, depending on the initial orbital period 
(Figs.~\ref{f:M_Porb_full_grid},~\ref{f:HR_1} and \ref{f:HR_2}.) 
This limit is slightly lower than $M_{1,0,{\rm min}}\approx7$\,\ms, obtained by 
%G\"{o}tberg et al. 
\cite{2018A&A...615A..78G}
for $q_0$=0.8 and a very early case B.

(iv) Stars more massive than $\approx$15\,\msun\  continue to expand after He ignition in nondegenerate cores and can still fill critical lobes. Later, the contraction that accompanies He burning terminates RLOF after a fraction of the H/He envelope
has been lost, but a relatively large fraction of it is still retained. 
Then the star may continue to burn He in the core close to the RG branch to become a red supergiant with a CO core. In some cases,
it may refill the Roche lobe.
But when the mass loss terminates and the star contracts, 
because of the presence of a relatively massive H/He envelope, it never becomes cool enough ($\log(\te) \leq 4.4$) to be considered in this study as a \hes\ star.
We looked into the evolution of such stars and found, as expected, that the remnants of these stars are massive enough 
to experience a SN explosion. Depending on the amount of H and He  retained in the envelope, despite the stellar wind and mass loss in the loops of their tracks in the HRD,
such SNe may be classified as Ib or IIb.

Thus, a fraction of the deemed 
progenitors of binaries with \hes\ stars may be lost. The progenitors of hot helium stars with masses between those of subdwarfs 
($\aplt$\,2\,\ms) and WR stars ($\apgt$\,7\,\ms) are not 
binaries with $M_1\apgt$\,2.5\,\msun, as is often claimed, but objects with $M_{1,0}$ at least 2.5 -- 3.5 times higher and with a limited range of ZAMS periods and mass ratios. This reduces their relative number compared to the subdwarfs and WR stars.

\section{Results of computations}
\label{sec:results}

\begin{figure} %2
 \resizebox{\hsize}{!}{\includegraphics{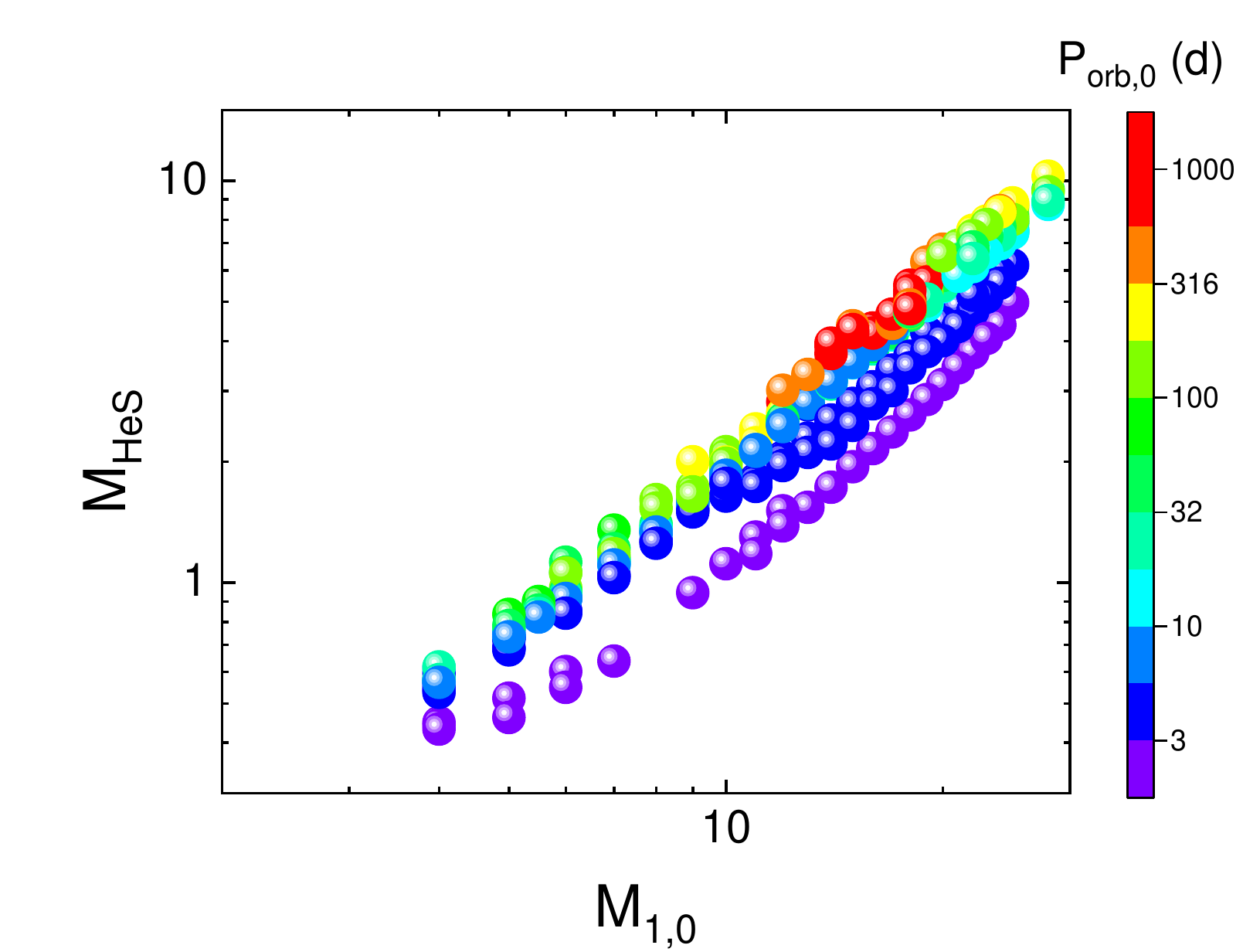}}
 \resizebox{\hsize}{!}{\includegraphics{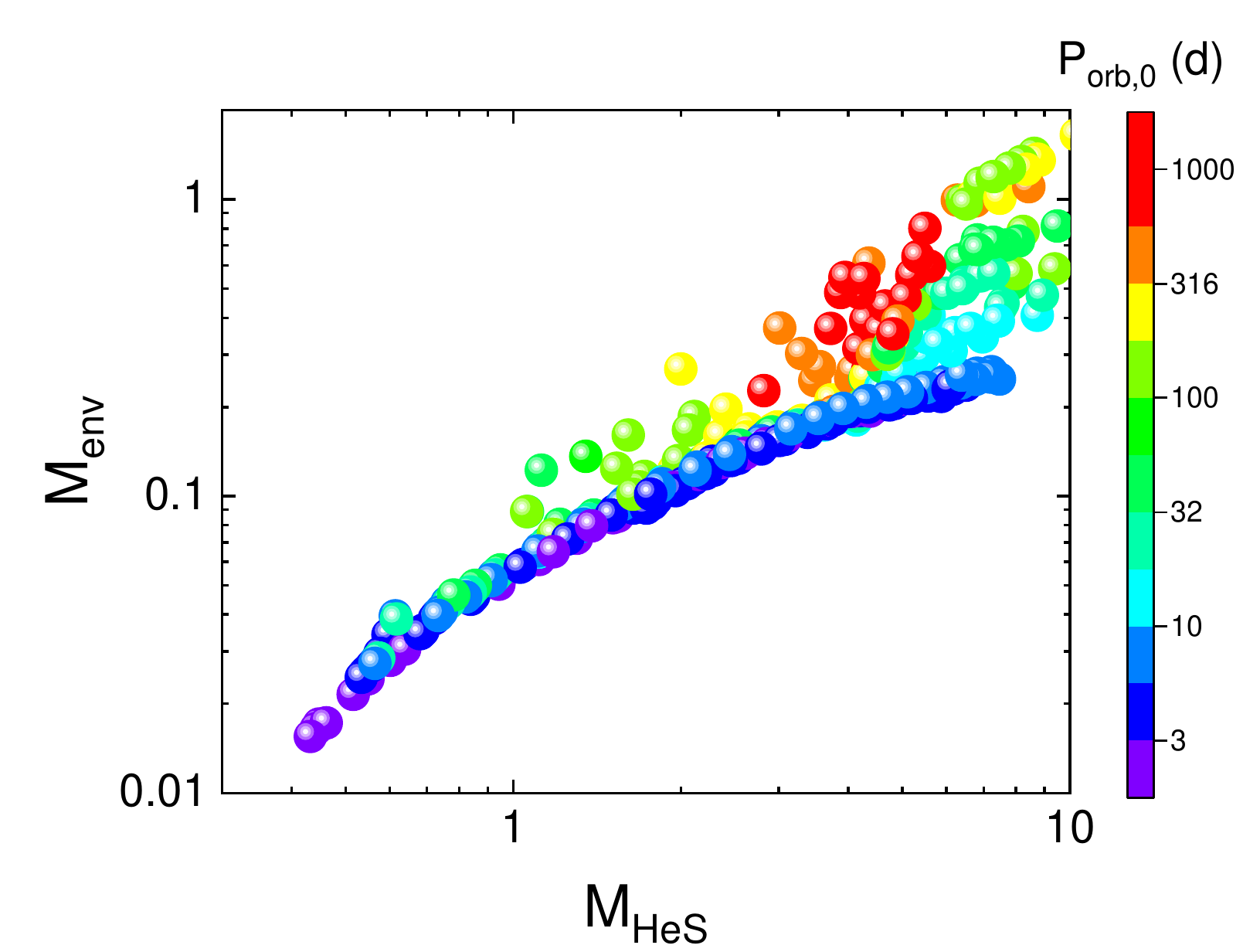}}
 \caption{ZAMS mass –- remnant mass ({\it upper panel}) and remnant mass –- H/He envelope mass ({\it lower panel}) relations for the HeS stars produced by the systems from the grid of initial systems.
Color-coded are ZAMS periods of the binaries. All masses are in solar units.}
\label{f:remn}
\end{figure}
%%%%%%%%%%%%%%%%%%%  FIG.3%   %%%%%%%%%
\begin{figure*}   %3
\centering
\includegraphics[width=17cm]{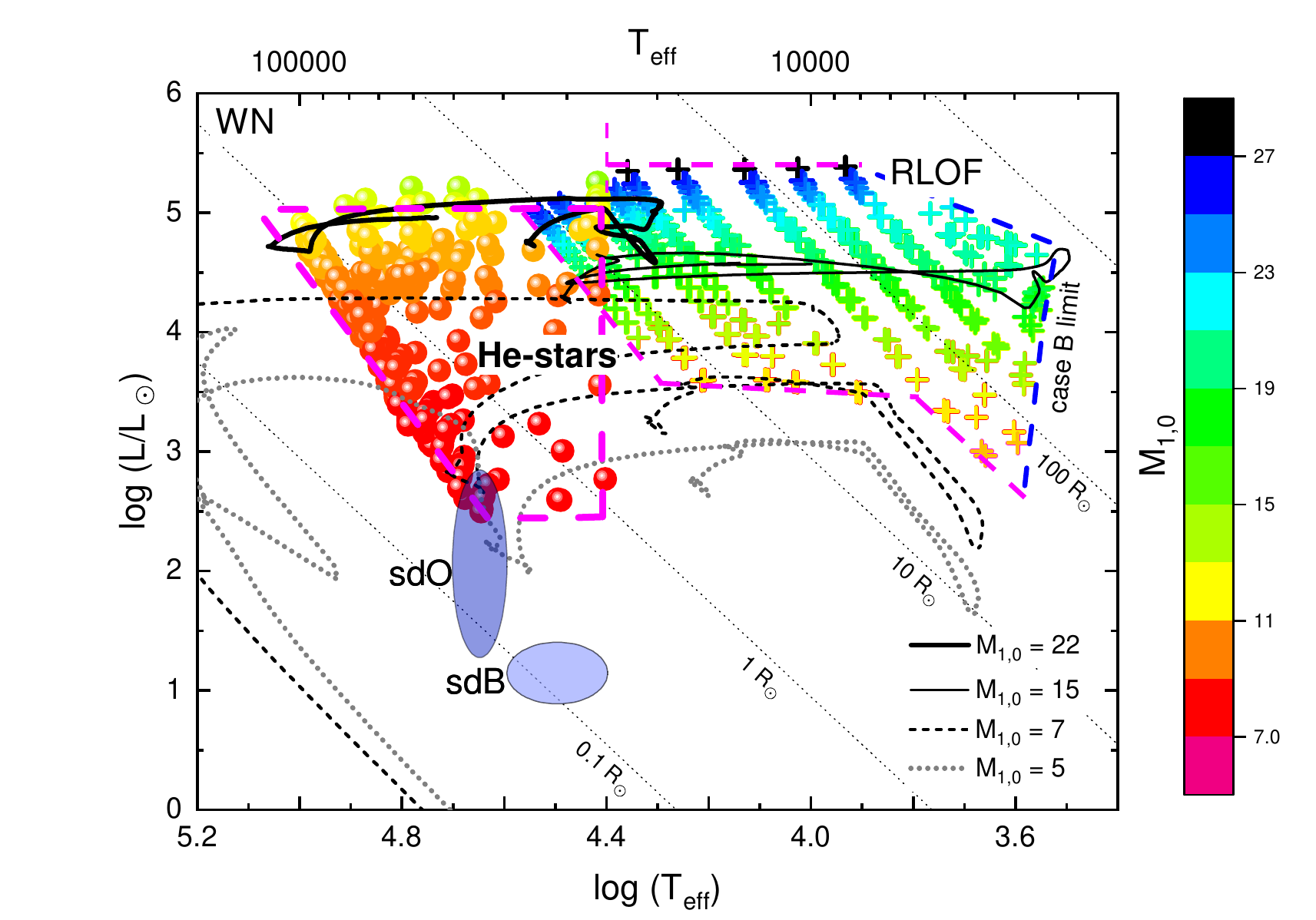}
\caption{Positions of \hes\ stars (at the minimum luminosity along their evolutionary tracks, see the text) and their progenitors in the HRD. The ZAMS masses are color-coded. The crosses mark the progenitors of \hes\ stars at the beginning of the RLOF, and the filled circles show  \hes\ stars descending from them. 
The progenitor's and descendant's symbols of the same color belong to the same evolutionary tracks,  i.e., red circles denote the descendants of 
red crosses, yellow circles the descendants of yellow crosses, and so on.
Dashed magenta lines bound the regions occupied by the Galactic WN stars, \hes\ stars, and precursors of the latter.
In the upper right corner, dashed blue lines indicate the limits of case B mass exchange and stable quasi-conservative mass exchange. The black symbols in the diagram indicate stars with ZAMS mass 28\,\msun, which produce WR stars. 
Thick and thin solid and dashed lines show the evolutionary tracks of the primaries of binaries with ZAMS masses of 22, 15, and 7,  and 5\,\ms, and initial periods of 5 (case A mass exchange), 550 (case B), 100 (case B), and 
50 days (case B), respectively, illustrating the formation paths of \hes\ stars. 
The thin dotted lines are the lines of equal stellar radii. The gray ellipses depict the locations of 
sdB and sdO stars \citep[from ][Fig.1]{2016PASP..128h2001H}.  
}
\label{f:HR_1}   
\end{figure*}
\begin{figure*} %fig 4
\centering
\includegraphics[width=17cm]{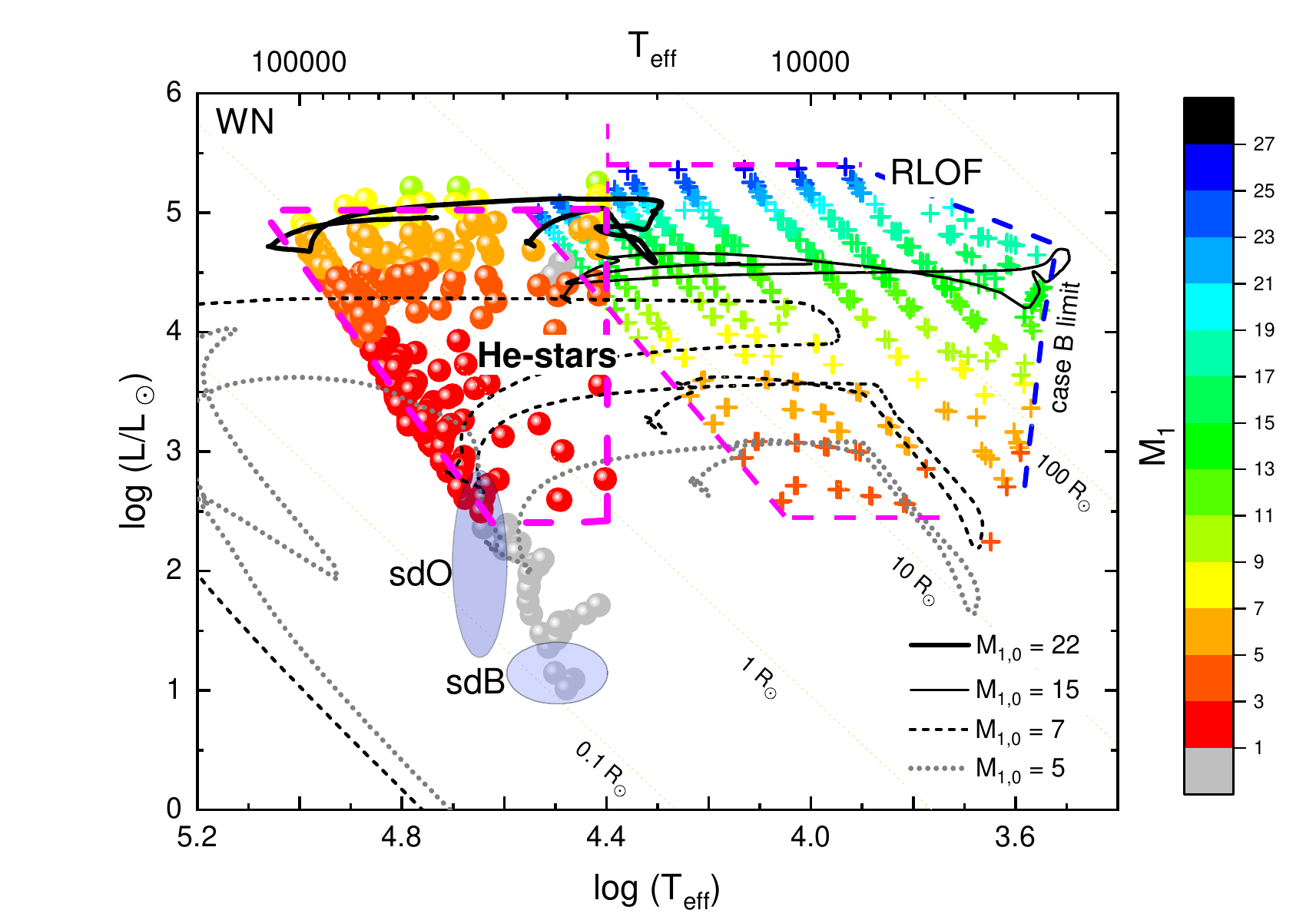}
\caption{ Same as in Fig.\,3, but the color scale
encodes the stellar masses at the RLOF and in the core He-burning stage after the RLOF. The entire stage of He core-burning stretches from the position of the circles by about $\pm 0.05$ in $\log(\te)$ and up by 
$\aplt$0.5 in $\log(L/\ls)$. 
Gray circles show the descendants of (5 -- 7)\,\ms\ ZAMS primaries producing HeS stars less massive than 1\,\ms\ in case B 
mass exchange. Initial parameters of the 7, 15, 22\,\ms\ tracks shown in the plot are the same as in Fig.~\ref{f:HR_1}. }
\label{f:HR_2}
\end{figure*}
%%%%%%%%%%%%%%%% END FIG. 4 %%%%%%
\subsection{Initial-final mass relation for progenitors of \hes\ stars and orbital periods of the binaries with \hes\ stars}

Figure~\ref{f:M_Porb_full_grid} shows
the full grid of computed models in the systems with initial mass ratios of components, $q_0=M_{2,0}/M_{1,0}$=0.6, 0.8, 0.9 in the diagram $M_{1,0}$ versus $P_0$.
For binaries with $q_0 {<} 0.6$, mass exchange is, as a rule, unstable. Nevertheless, some systems with $q_0=0.4$ avoid components merging and produce 2\,\ms\ to 7\,\ms\ \hes\ stars with 3.5\,\ms\ to 7.5\,\ms\ companions (see Fig.~\ref{f:fig15_4_2log} below.)
 
The masses of stripped stars for the analysis are taken at the positions along the tracks where the luminosity  reaches its minimum. Close to these points, HeS stars spend most of the core He-burning time, which is $\sim10\%$ of the main-sequence lifetime. Figure ~\ref{f:M_Porb_full_grid} also shows the ``fate'' of the binaries.  
 
 The positions of colored regions in the left panel of  Fig.~\ref{f:M_Porb_full_grid} illustrate the effect of the 
``guillotine'' factors that define which ZAMS binaries may produce \hes\ stars via
stable nonconservative mass exchange, as discussed above in Sec.\,2.3. The systems with masses of the H/He envelopes of remnants exceeding 0.3\,\ms\ (open circles) are highlighted because, if the remnants retain such a mass of the envelope after the He-burning stage, they do not expand after the core He-exhaustion and do not lose mass due to the refilling of critical lobes.  

The masses of envelopes, as well as the masses of \hes\ stars, are to some extent uncertain, because after cessation of RLOF they decrease by the stellar wind as stars evolve toward higher \te. Masses of envelopes (and masses of \hes\ stars) in the core He-burning stage may remain almost the same if \citet{2017A&A...607L...8V} stellar wind mass loss is used, or decrease by several 0.1\,\ms\ if \citet{2000A&A...360..227N}'s law is applied. The figure shows that the star may become really ``naked,'' virtually without any hydrogen at the surface.  Since we applied Nugis \& Lamers' recipe, this possibility is quite realistic (see also Figs.~\ref{f:remn}, \ref{f:a1}, and \ref{f:a2} in the appendix.) However, it should be noted that the issue of a ``correct'' mass-loss rate law is not solved as of yet.

Figure~\ref{f:remn} shows ``ZAMS mass -- remnant mass'' and ``remnant mass -- envelope mass'' relations for the HeS stars produced by the systems from the grid of initial systems. Like in Fig.~\ref{f:M_Porb_full_grid}, the masses of \hes\ stars and their envelopes are shown for the lowest luminosity point along the evolutionary track, where helium stars spend most of their lifetime. A feature seen both in Figs.~\ref{f:M_Porb_full_grid} and \ref{f:remn} is a quite weak dependence of \hes\ star masses on $\porb_{,0}$. The reason for this is that there is only a small change in He-core masses 
during the rapid crossing of the Hertzsprung gap. Stars with 
$M_{\rm HeS}\geq$\,2\ms\ in the stable nonconservative mass-exchange channel are produced only by binaries with 
$P_{\rm orb,0} \apgt 10$\,days and masses $\gtrsim(10 - 12)$\,\ms. Moreover, 
it is clear that for $M_{\rm HeS,min}=$\,1\ms\
the lower limit of the progenitor masses  is close to 6\,\ms, while the initial orbital periods should exceed two to three days.   
However, the remnants of stars with $M_{1,0}$ slightly below 6\,\ms\ may also spend some short time in the \hes\ stars'  domain of the HRD during the shell helium burning, as illustrated below by the track for a star with a ZAMS mass of 5\,\ms\ in Fig.~\ref{f:HR_2} (the lower dotted track.) 

Generally speaking, the outcome of the evolution and masses of \hes\ stars rather weakly depends on the initial mass ratio $q_0$. This is related to a very weak dependence of the radii of critical lobes on $q_0$. For a given $M_{1,0}$ and 
$P_{orb,0}$,  
$R_{\rm cr} \propto (1+q)^{-1/3}$. Noticeable as well is a slightly steeper increase in masses of \hes\ stars with an increase in $M_{1,0}$, as $q_0$ decreases. 

\subsection{Synthesized \hes\ stars and their progenitors in the HRD}

In Fig. \ref{f:HR_1} we present the relative positions of some of the computed \hes\ stars and their progenitors in the HRD.
We mark with similar colors the progenitors  of \hes\ stars (crosses) and their descendants (circles) in order to show their relative displacement in the HRD. In this figure and in Fig.~\ref{f:HR_2}, the domains 
occupied by \hes\ stars and their progenitors and by the Galactic WN stars according to 
\citet{2020A&A...634A..79S} (in the upper left corner of the plot) are outlined 
with dashed magenta lines.

In the low \te\ part of the HRD, the domain of progenitors for stars less massive than 15\,\ms\ is limited by the core He ignition 
line, since the latter causes overall contraction of stars. 
For more massive stars, the low \te\ limit is due to high mass-loss rates resulting in the formation of common envelopes or the formation of mass-loss remnants with relatively massive H/He envelopes that contract but that do not reach the \te=25\,000\,K required to
classify them as \hes\ stars, as was explained above. Additionally, there are stars in which RLOF occurs close to the Hayashi line but is terminated after only a part of the envelope is lost, because He burning becomes the dominant energy source. These stars continue their evolution as red supergiants.  

Figure \ref{f:HR_1} shows some of the computed models of \hes\ stars at the positions where their luminosity along the tracks reaches its minimum.
Close to these points, \hes\ stars spend a fraction of the core He-burning time when they have the core He abundance  $\simeq$0.5. The rest of the core He burning occurs when the stars evolve to higher luminosities and \te. In the latter stage, \te\ may increase by
$\Delta(\log(\te)) \approx 0.05$, while $\Delta(\log(L/\ls))$ may be 
up to 0.5, as is seen
for the tracks plotted in Figs.~\ref{f:HR_1} and \ref{f:HR_2}. The time spent in the ``ascending'' branch of the track is comparable to the time spent around the luminosity minimum.  These core He-burning stars create a
subpopulation of \hes\ stars on the hot side of the strip of stars with minimum luminosity during the core-He burning stage, which is clearly seen below in Fig.~\ref{f:HR2_n10}.  

In addition, we plot in the diagram the track of an $M \simeq 0.7$\,\ms\ remnant of a 5\,\ms\ star (the lower blue line) that enters the domain of \hes\ stars only for a very short time in the stage of contraction, when an H shell still dominates in luminosity and during He shell burning (the part of the track turning up at $\log(\te)\approx$4.6\ and then leftward.) The He shell burning is unstable, and the track makes loops in the HRD, a part of which extends beyond the left limit of the plot. 
  
In this figure, the track of a 22\,\ms\ primary star in the system with $q_0$=0.6 is a kind of ``limiting'' one for the systems experiencing case A mass exchange. The core He burning of more massive stars occurs in the WR-stars domain of the HRD. For case B, this limiting mass is 24\,\ms. 

Figure \ref{f:HR_2} displays the positions of  \hes\ 
stars  and their progenitors in the HRD. 
It is clear that the population of \hes\ stars
is  distinct from the population of ``canonical'' subdwarfs and is bridging locations of sdO and WR stars in the HRD.
This is in agreement with the results  of \citet{2018A&A...615A..78G}, which suggest that subdwarfs, \hes, and WR stars form a continuous spectral sequence
where the strength of the He\,{\sc ii}$ \lambda 4686$\,\AA\ line increases from absorption to emission.

Both Figs.~\ref{f:HR_1} and \ref{f:HR_2}
suggest that the population of \hes\ stars is dominated by low-mass objects ($\aplt$4\,\ms) for a wide range of ZAMS masses up to $\approx$15\,\ms.  This reflects the initial mass function of the primary components of binaries and variation in the range of He-core masses.
It is noteworthy to mention that in the HRD the domain of the most massive \hes\ stars
($\mhe\apgt$5\,\ms) overlaps with the location of 
the main-sequence stars more massive than $\sim 10$\,\ms. There is a kind 
of gap in the HRD at $\log(\te) \approx (4.0 - 4.4)$ and $\Delta(\log(L/\ls)) \approx (2.5 - 4.2)$.  It is populated by the stars that are not hot enough to be classified as \hes\ stars but that have He-enriched envelopes. 

Figures~\ref{f:HR_1} and \ref{f:HR_2} 
indicate that the formation of massive sdO stars (1 -- 2)\,\ms\ may be explained if they descend from the stars with ZAMS masses $\approx$(5 -- 10)\,\msun, in contrast to  the canonical $\sim$0.5\,\ms\ subdwarfs with progenitor masses below 2\,\ms\ 
\citep{1985ApJS...58..661I,2002MNRAS.336..449H,2003MNRAS.341..669H,2005ARep...49..871Y}. Interestingly,
Fig.~\ref{f:HR_2} shows that known massive sdO companions of Be-stars \citep{2023AJ....165..203W} may be either in the core or shell He-burning stage.\\

\subsection{Synthesized Galactic \hes\ population properties }

%%%%%%%%%%%%%%%%%%% FIG 5 %%%%%%%%%%%%%%%%%
\begin{figure}    %5 version 24.07.2023
%\centering
%\includegraphics[width=8.4cm]{fig5a.pdf}
%\hskip -0.089 cm
%\includegraphics[width=8.4cm]{fig5b.pdf} 
 \resizebox{\hsize}{!}{\includegraphics{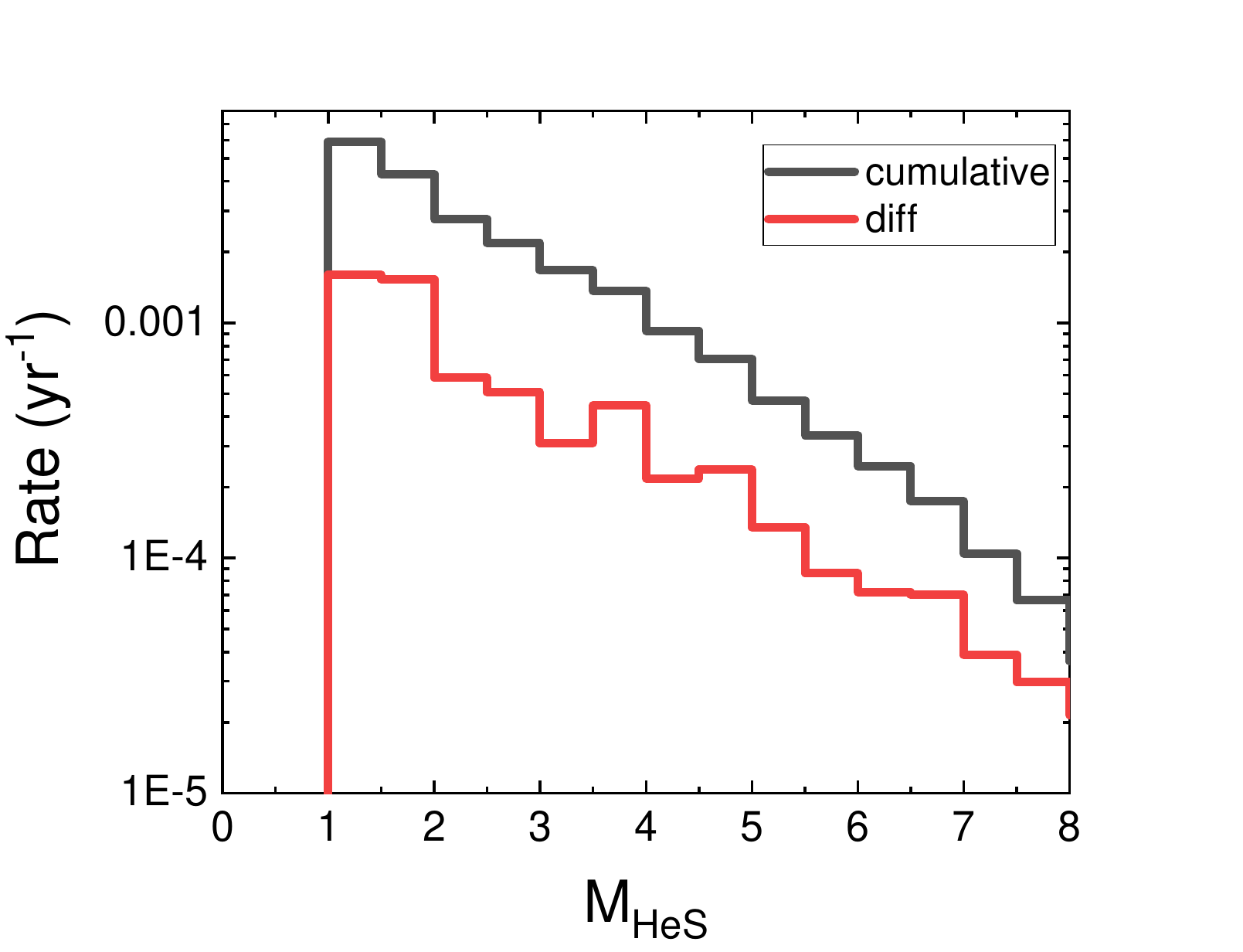}}
  \resizebox{\hsize}{!}{\includegraphics{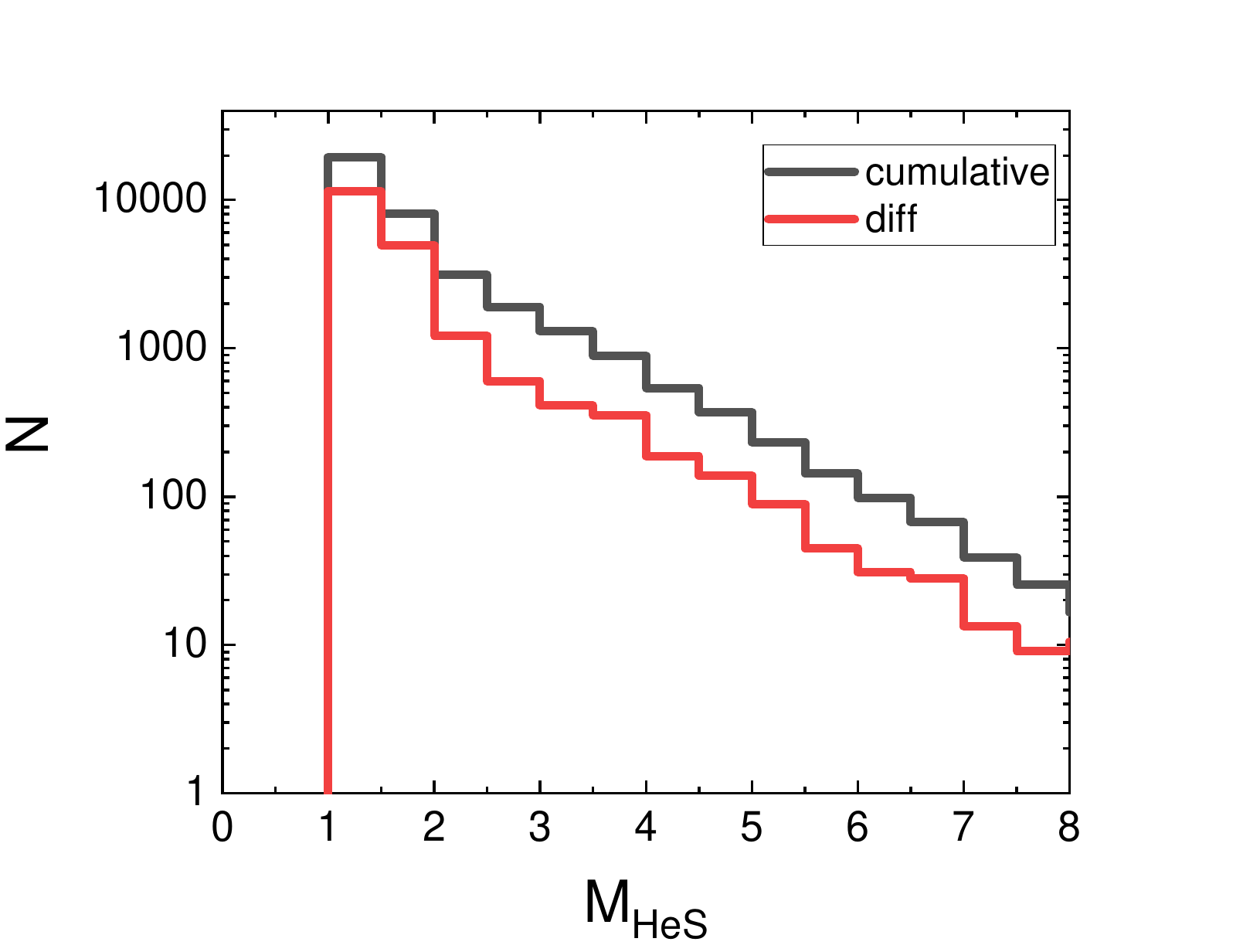}}
\caption{Differential and cumulative distributions of the formation rate of 
\hes\ stars with masses $\geq 1$\,\ms\ as a function of  the stripped star mass 
{\it (upper panel)}.  
Differential and cumulative distributions of the total number of HeS stars in the Galaxy {\it (lower panel)}.   }
\label{m1_distr} 
\end{figure}
%%%%%%%%%%%%%%%  END FIG 5 %%%%%%%%%%%%%%

Figure~\ref{m1_distr} shows differential and cumulative distributions of the formation rate and the total number of binaries containing \hes\ stars. 
For the limiting mass of \hes\ stars, $M_{\rm lim}$ = 1\,\ms, their formation rate is only about 1/170\,\pyr\ and their number in the Galaxy is close to 19\,500.
If $M_{\rm lim}$ = 2\,\ms, the formation rate and number of \hes\ stars sharply decrease to about 1/360\,\pyr\ and 
3\,100, respectively. The dominance of low-mass \hes\ stars, actually overlapping with the mass range of the most massive sdO stars, is clearly seen. 

%%%%%%%%%%%%%%%%%%% Fig 6 %%%%%%%%%%%%%%%%%%%%%%%%
\begin{figure}    %6 version 28.10.23 
 \resizebox{\hsize}{!}{\includegraphics{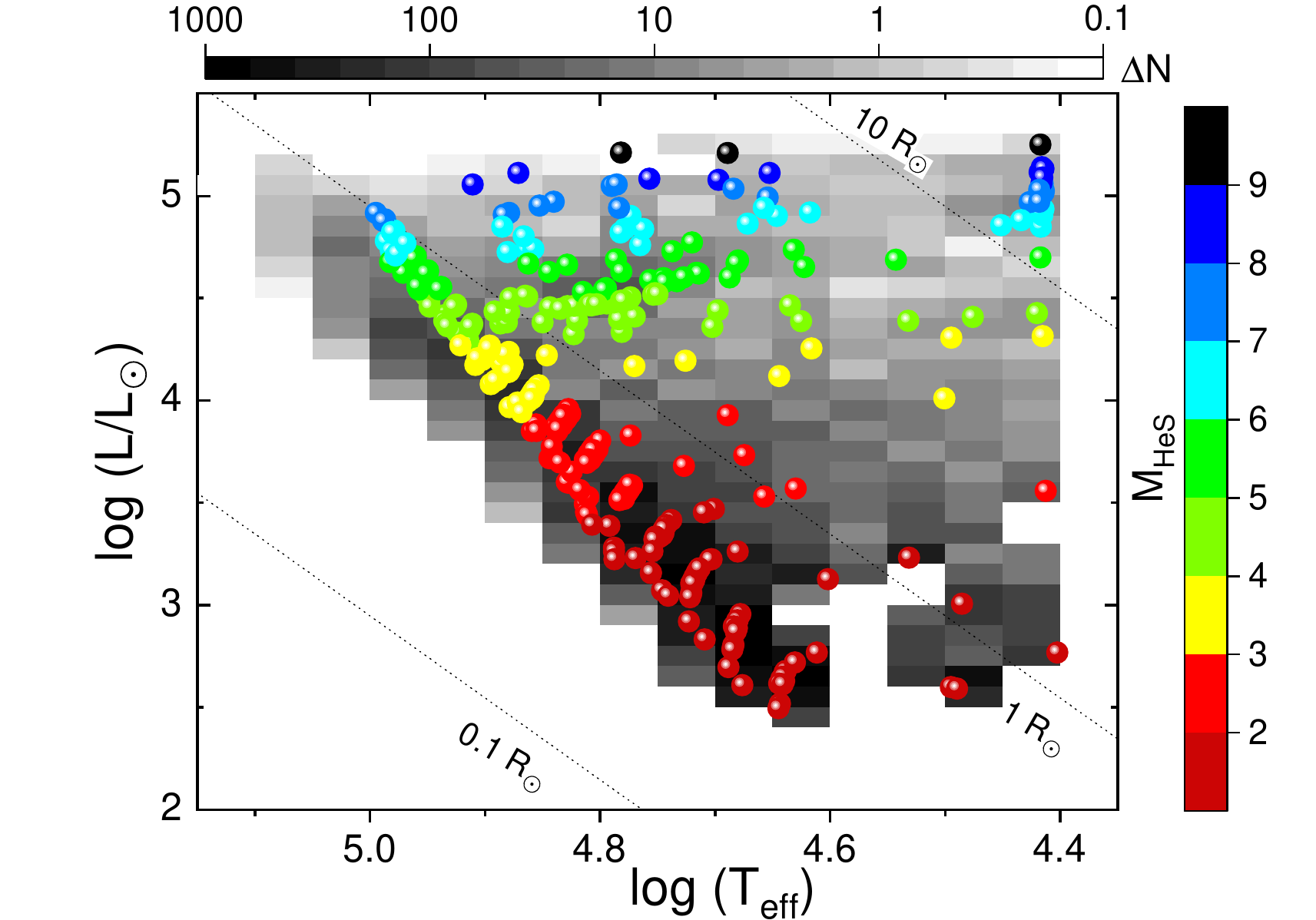}}
 \resizebox{\hsize}{!}{\includegraphics{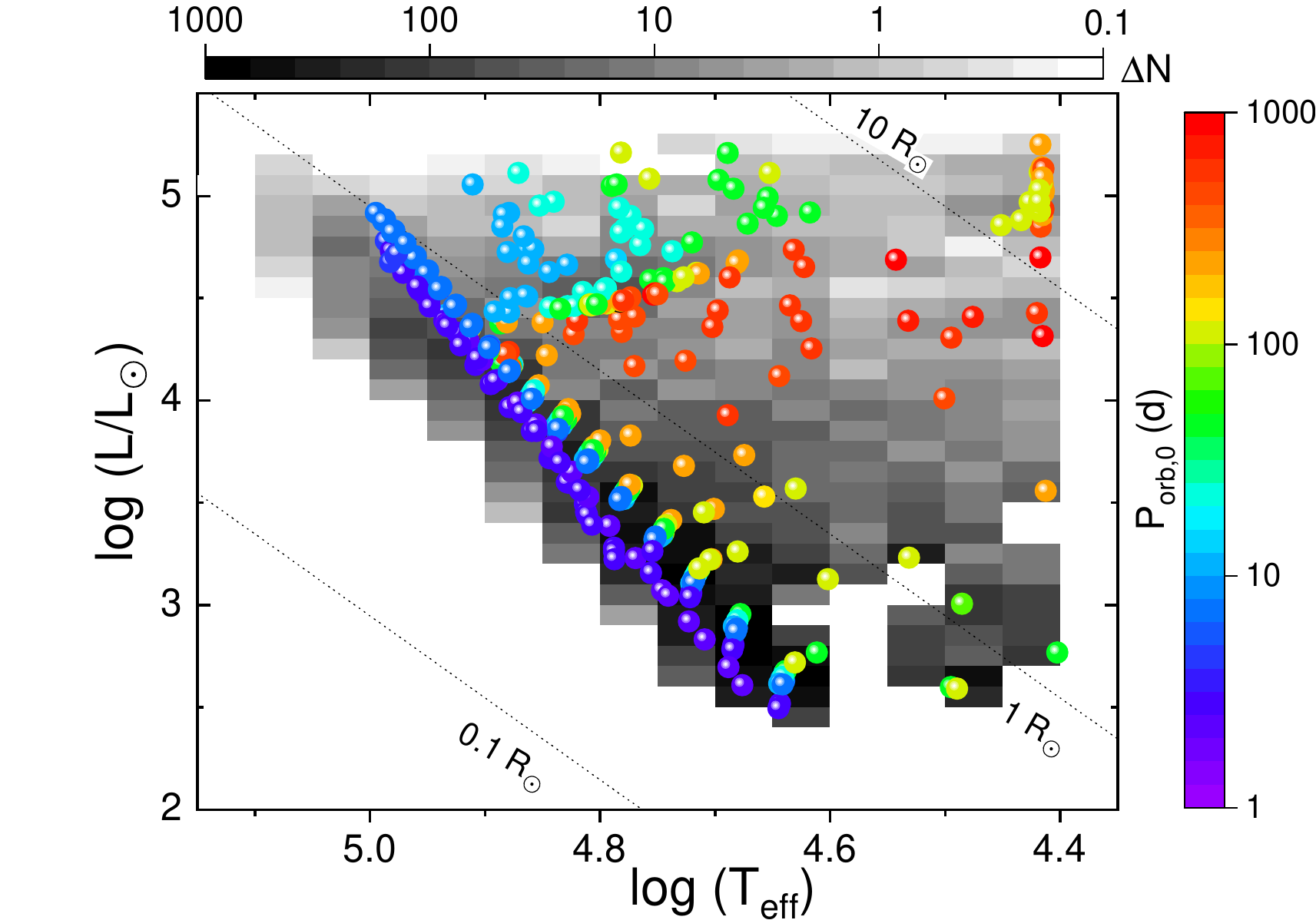}} 
\caption{HRD of the synthesized population of \hes\ stars. The gray scale shows the number of \hes\ stars per  $\Delta(\log(\te))\times\Delta(\log(L/\ls))$ = 0.05$\times$0.1 pixels.
{\it Upper panel:} Symbols are color-coded according to the \hes\ stars' masses (scale to the right.) {\it Lower panel:} Symbols are color-coded according to the initial binary periods.  Circles mark the positions of the computed models, like in Figs.\,3 and 4.    }
\label{f:HR2_n10}
\end{figure}
%%%%%%%%%%%%%%%%%%%%%%% END Fig 6 %%%%%%%%%%%%%%%%%%%%%

%%%%%%%%%%%%%%%%%  FIG 7 %%%%%%%%%%%%%%%%%%%%%%
\begin{figure}   %7
\resizebox{\hsize}{!}{\includegraphics{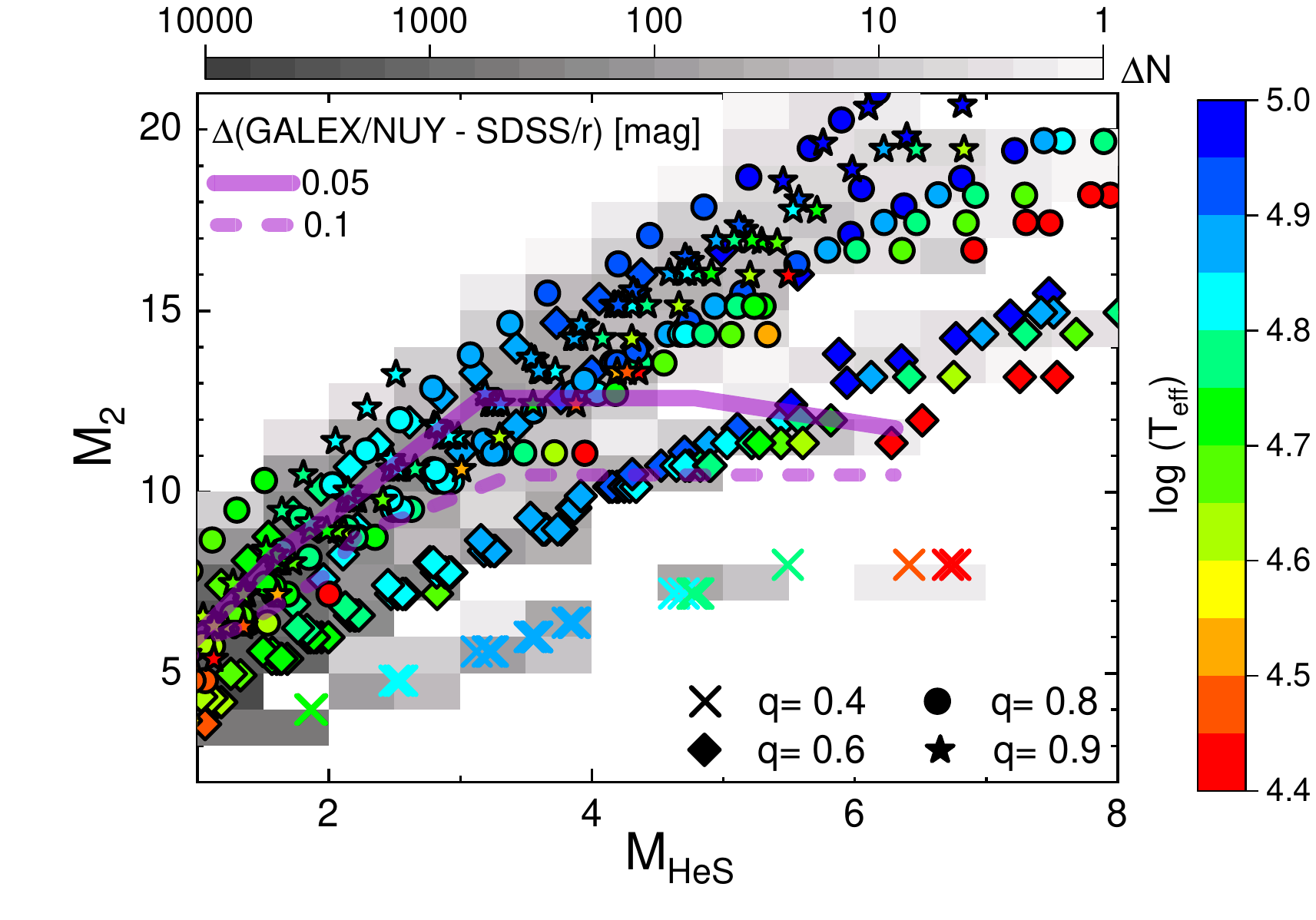}}
\caption{Relation between masses of components in the synthesized population of binaries with \hes\ stars.
Different symbols denote the descendants of binaries with different initial mass ratios of components, $q_0$. 
The symbols are color-coded by the \te\ of \hes\ stars.
The gray scale codes the number of stars per 
$\Delta(M_{\rm HeS})\times\Delta(M_2)$ = 1.0\,\ms$\times$1.0\,\ms\ pixel.
Solid and dashed magenta lines show roughly two limiting values of the UV color excess 
{\it (GALEX/NUV-SDSS/r)} computed by \citet{2018A&A...615A..78G}, below which \hes\ stars may be detected due to their color excess compared to the UV color of the companion (assumed to be a main-sequence star.) } 
\label{f:fig15_4_2log}
\end{figure}
%%%%%%%%%%%%%%%%%%  END FIG 7 %%%%%%%%%%%%%%%%%%

%%%%%%%%%%%%%%  Fig 8 %%%%%%%%%%%%%%%%%%%%%%
\begin{figure}
\resizebox{\hsize}{!}{\includegraphics{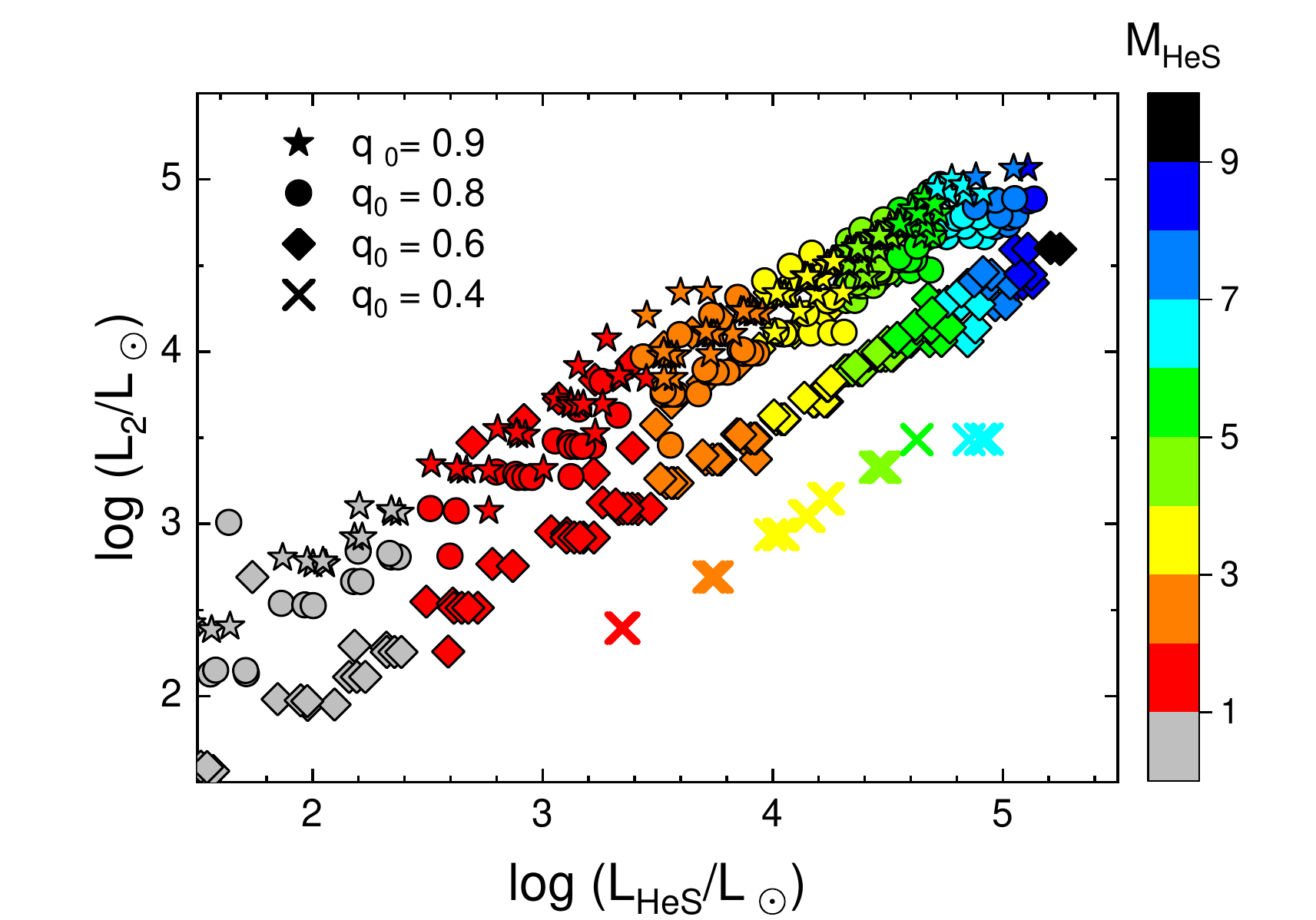}}
\caption{Relation between luminosities of \hes\ stars and their companions in the computed binaries with different initial mass ratios (symbols in the insert.) The gray symbols show systems that failed to produce \hes\ stars. The color scale codes the masses of \hes\ stars. }
\label{f:lhe_l2}
\end{figure}
%%%%%%%%%%%%%%%%%%%%%%%% END Fig 8 

%%%%%%%%%%%%%%%%%%%%%%%%%%  Fig 9 %%%%%%%%%%%%%%%%%%%%
\begin{figure}    
\resizebox{\hsize}{!}{\includegraphics{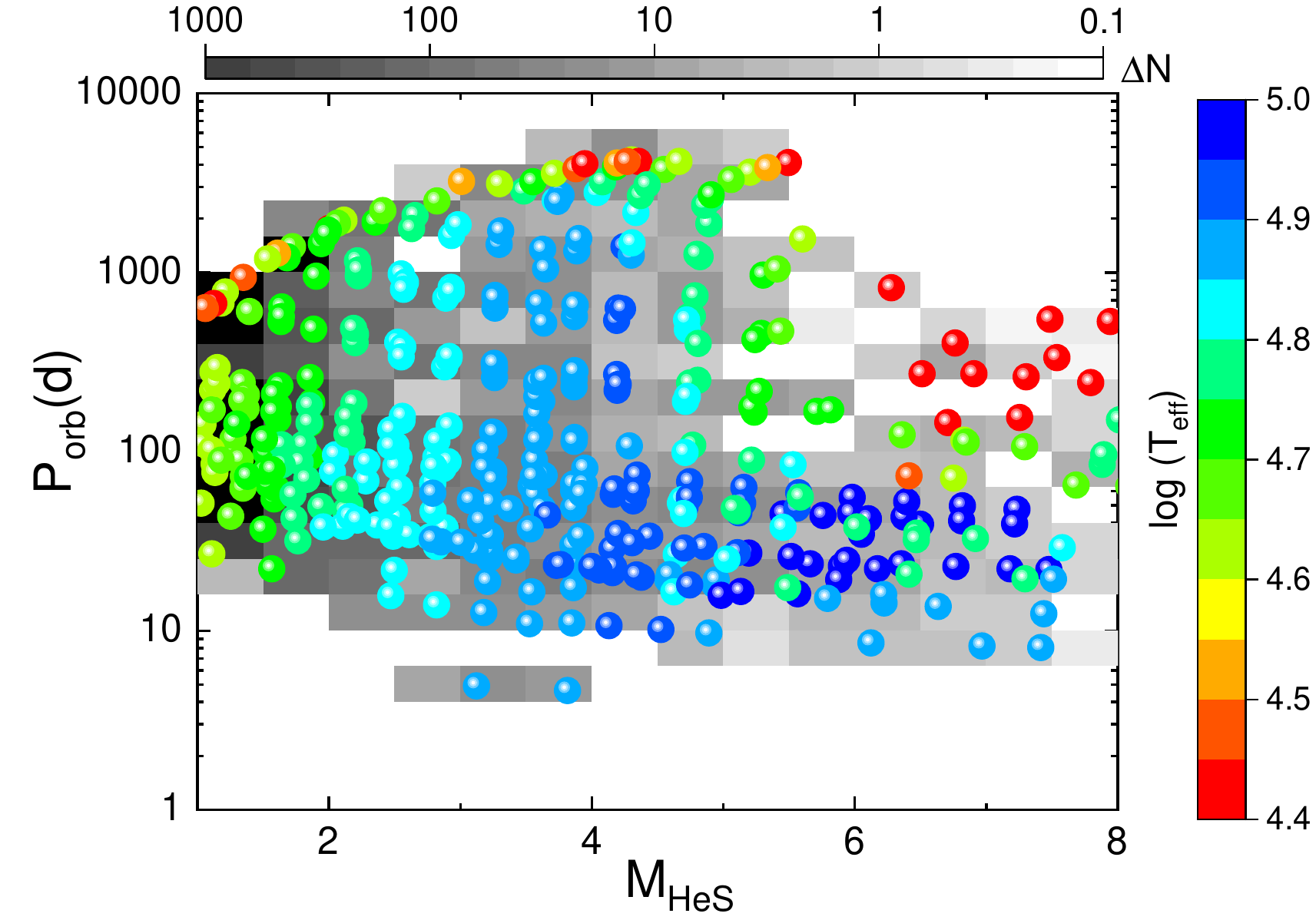}}
\caption{Relation between HeS star masses and the orbital periods of binaries harboring them. The gray scale codes the Galactic number of stars per pixel.
The color scale codes the effective temperatures of the stars. 
}
\label{m1_porb} 
\end{figure}
%%%%%%%%%%%%%%%%%  END FIG  9 %%%%%%%%%%%%%%%% 

%%%%%%%%%%%%%%%%%% FIG 10 %%%%%%%%%%%%%%%%%%%%%
\begin{figure*}  %10
\centering
\includegraphics[width=8.4cm]{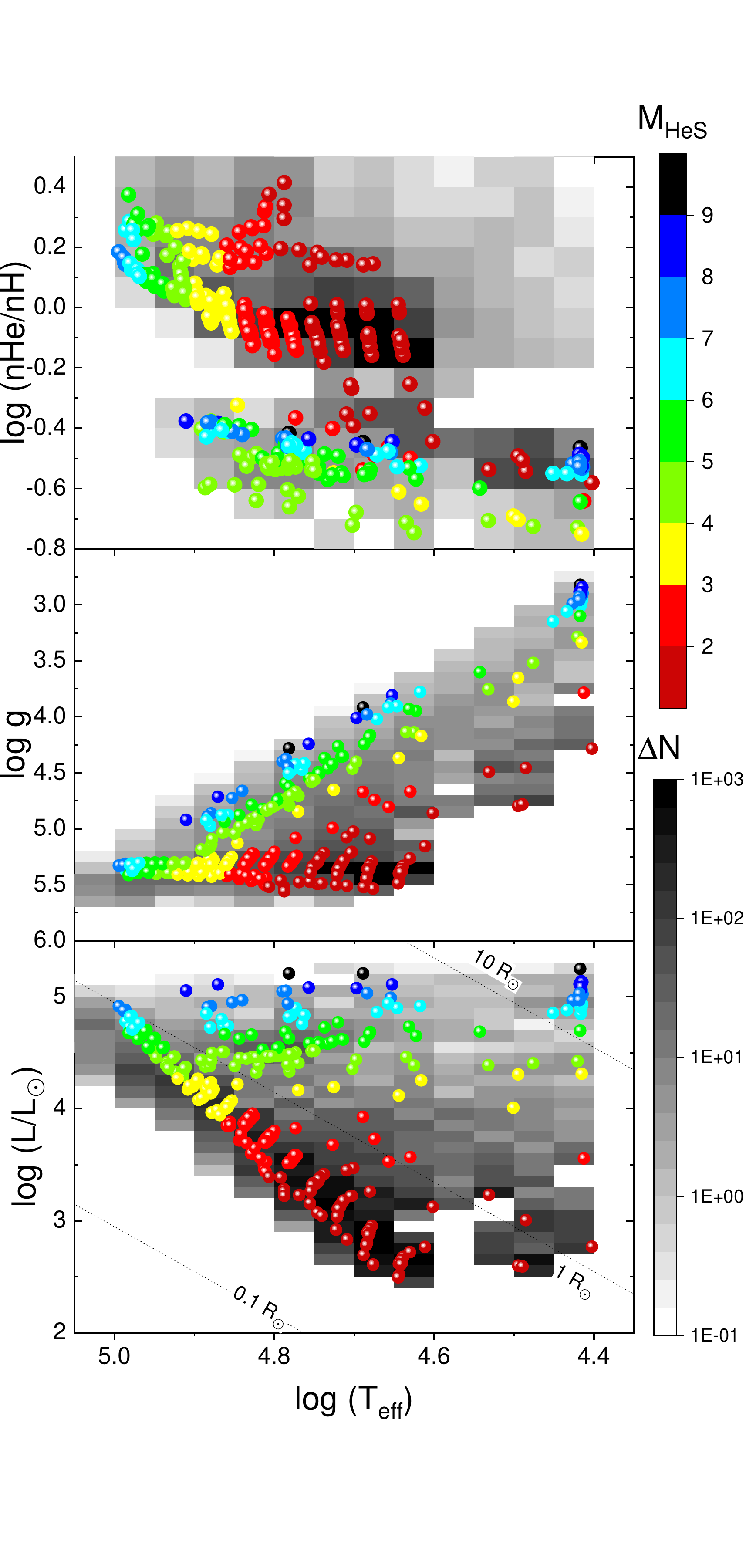}
\hskip -0.089 cm
\includegraphics[width=8.4cm]{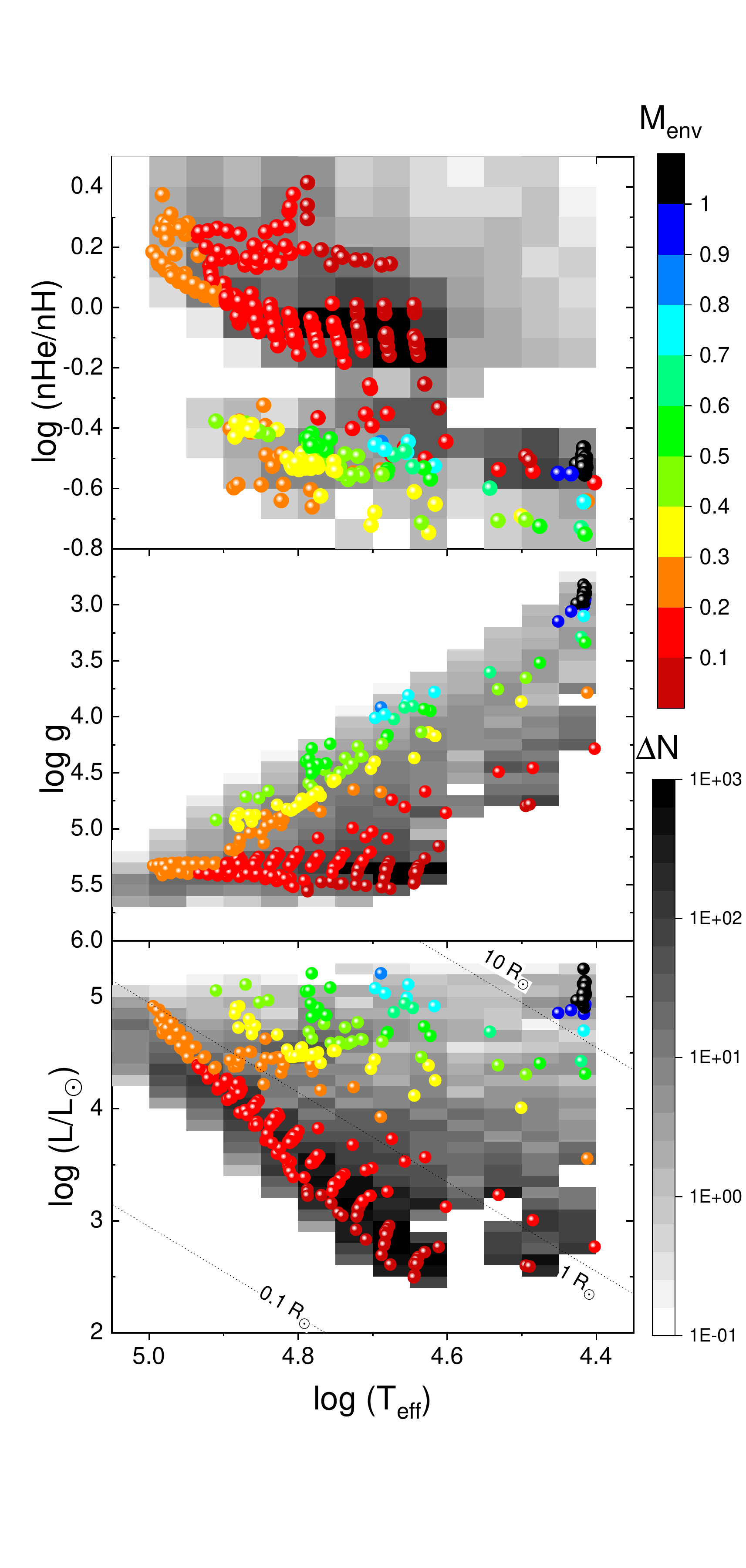}
\caption{Parameters of the synthesized population of \hes\ stars. {\it Upper panel:} Distribution of the surface He-abundance ${\rm (n{He}/n{H})}$; {\it middle panel:} Log of surface gravity ($\log(g)$); {\it lower panel:} HRD.  The symbols are the same as in 
Fig.\,6.
The gray scales in all panels show the Galactic number of systems per pixel. The colors in the left panels encode the masses of \hes\ stars, while the colors in the right panels encode the masses of H/He envelopes.}  
\label{f:He_H_over}
\end{figure*}
%%%%%%%%%%%%%%%  END FIG 10 %%%%%%%%%%%%%%%%%%%%

The HRD of the synthesized population of \hes\ stars is shown in Fig.\,\ref{f:HR2_n10}. We note the absence of colored circles in the leftmost part of the shaded region. This is because we plot the positions of \hes\ stars in the HRD at the minimum luminosity, in other words at an evolutionary stage where the stars spend a substantial fraction of their core He-burning time.
However, a total exhaustion of He in the core occurs at a slightly higher luminosity (see captions to Figs.~\ref{f:HR_1} and \ref{f:HR_2}.) 

For the stars that make loops in the HRD in the He shell burning stage during which \logte\ becomes lower than 4.4, we took into account the time spent by them before  \logte\ becomes lower than 4.4 for the first time. The remaining He shell burning time is short and can be safely neglected.

As can be seen in 
Fig.\,\ref{f:HR2_n10}, the majority of \hes\ stars populate a rather narrow ($\Delta(\log(\te)) \approx$0.2) ``strip''  between $\log(\te)\approx$4.6, $\log(L/\ls)$=2.5 and 
$\log(\te)\approx$4.9, $\log(L/\ls)$=4. The objects located at a lower \te\  originate in binaries with relatively large ($\apgt 100$\,days) initial orbital periods, which retain massive H/He envelopes and burn the core He at  $\te {<} 25\,000$\,K, in other words below the limit we adopted for \hes\ stars.  

There are relatively densely populated branches of 
$\simeq$5\,\ms\ \hes\ stars extending to a lower \te\ from the main ``strip.'' These are the descendants of binary components more massive than about 15\,\ms\ that retained relatively massive H/He envelopes, but not exceeding 0.3\,\ms. The stars with more massive envelopes never reach 
$\log(\te)=4.4$. We distinguish the latter as a separate category of stars with He-enriched envelopes, as already mentioned in Sec.\,3.3.

The most luminous among our synthetic population are WR stars (these are marked in black and three dark shades of blue in the upper panel of Fig.\,\ref{f:HR2_n10}.) 
The position of the most massive \hes\ stars overlaps with that of WR stars at
 $\logte \approx$(4.7 -- 4.9) because \hes\ stars spend a short shell helium-burning time at high luminosities. 
 But it is evident that such stars should be scarce. 
It is worth noticing that the remnants of case A mass exchange systems belong to the 
 population with the highest \te\ and always burn He close to the minimum luminosity point of the track (see the lower panel in Fig.\,\ref{f:HR2_n10}.) 

Figure \ref{f:fig15_4_2log}  shows the masses of components of binaries in the synthesized population as a function of the initial mass ratio of components. 
The effective temperatures of stars have a weak trend, visible already in Fig.~\ref{f:HR_2}: the most numerous \hes\ stars with masses below about 3\,\ms\ have the highest \te. In addition, the trend of the \te\ to decline with an increase in the \hes\ star mass is seen.

According to our assumptions about the mass and angular momentum loss (see Sec.\,2.1), the mass of the 
accretor changes only by several percent. Therefore, the minimum masses of companions to \hes\ stars are very close to those in the model grid, that is, \ 4\,\ms, which corresponds to the spectral type B7V ($\log(L/\ls) \approx 2.5$, \te=14\,000\,K, \citet{2013ApJS..208....9P}.)

It is suggested that \hes\ stars may be detected thanks to the excess of the UV emission of the binary hosting a \hes\ star compared to the UV emission of a single star with a mass equal to the main-sequence mass of its companion \citep{2018A&A...615A..78G}. We roughly plot in Fig.~\ref{f:fig15_4_2log} the lines for two values of  
color excess  {\it (GALEX/NUV-SDSS/r)}, as computed by G\"{o}tberg et al. Helium stars with companions that have masses below the limits 
shown by these lines may be detectable. 

Figure~\ref{f:fig15_4_2log} suggests that for the color excess $\aplt 0.05$, in the absence of other selection effects, about 50\%  
($\simeq1\,500$) of the binaries hosting \hes\ stars 
more massive than  $\simeq 2$\,\ms\ (approximately, the mass of the most massive known sdO) are ``detectable.'' The other 50\% are outshone in UV by their companions.
As was shown by G\"{o}tberg et al., using the excess in the emission line He~{\sc ii}~$\lambda$4686 as a signature of the presence of  a \hes\ star provides similar results. By no means is the number $\simeq1500$ quoted above an upper limit for a potentially detectable population of \hes\ stars.  Clearly, more dedicated work is needed to identify potential ``smoking gun'' spectral and photometric features of \hes\ stars in binaries. Such a study will be presented in a follow-up work.   

Figure~\ref{f:fig15_4_2log} also demonstrates several effects related to the initial mass ratio, $q_0$. First, the range of $\rm{M_{He}}$ only slightly depends on the initial $q_0$; this is a consequence of the weak dependence of the radii of the critical lobes on $q$. Second, it confirms an increase in the steepness 
of the mass ratio of components in the systems harboring \hes\ stars ($\rm{M_{He}/M_2}$) with an increase in the initial $q_0$, as was seen already in Fig.~\ref{f:M_Porb_full_grid}. This is a direct result of the algorithm for mass and angular momentum loss from the system, which strongly limits the amount of accreted mass by several
percent of the initial accretor's mass. Finally, it shows that, as long as $q_0 \apgt 0.6$, both cases A and B of mass exchange enable the formation of \hes\ stars, within the primary ZAMS mass 
and orbital period limits outlined in Figs.~\ref{f:M_Porb_full_grid},\ref{f:HR_1}, \ref{f:HR_2}. 
Figure~\ref{f:fig15_4_2log}  suggests that a significant fraction of \hes\ stars with masses (2 -- 4)\,\ms\ are potentially detectable.  

The relation between the luminosities of \hes\ stars 
(at the minimum luminosity in the core He-burning stage, $L_{\rm He}$, as in other figures) and the luminosities of their companions, $L_{\rm 2}$, at the same instant of time is shown in Fig.~\ref{f:lhe_l2}. As can be seen, $L_{\rm He}$ may be both lower or higher than $L_{\rm 2}$ depending on the \hes\ star's mass. 
Considering that bolometric corrections will favor the optical luminosity of main-sequence stars, one may expect that the majority of \hes\ stars will be outshone by their companions in the visual, while in UV the situation may be more favorable. But this could be confirmed only by computing
bolometric corrections for both components. Since in the shell He-burning stage the luminosity of helium stars is higher than in the core He-burning stage, the former may be more favorable for the detection of \hes\ stars; however, its short duration acts in the opposite direction.
We also show in this figure the relation $L_{\rm He} - L_{\rm 2}$ for several systems descending from binaries with 
$M_{\rm 1,0} = (5 - 7)$\,\ms, but not producing \hes\ stars more massive than 1\,\ms\ (the gray symbols in Fig.~\ref{f:HR_2}.)

Figure~\ref{m1_porb} shows  the relation between the masses of \hes\ stars and the orbital periods of the binaries harboring them, overplotted over the distribution of the synthesized population. Colors code the $\log(\te)$ of \hes\ stars. We note the absence of systems with $\porb \aplt$\,4 days and larger than 5\,000\,days. 
Most of the binaries have orbital periods ranging from  10 to 1\,000 days. 
The orbital periods do not correlate with \te. Both ``hot'' and ``cold'' \hes\ stars populate the same range of \porb.
The figure confirms the conclusions made before: the most massive \hes\ stars often have a low \te\ (within our assigned limit, $\log(\te) \geq 4.4$, for  \hes\ stars) due to the presence of massive H/He envelopes. 
Remarkably, the range of \porb\ of most systems is the same as the range of periods of their progenitor systems, below about 1\,000\,days, see
Fig.~\ref{f:M_Porb_full_grid}.

Figure~\ref{m1_porb}  clearly suggests that a large \porb\ is one of the factors 
hampering the discovery of \hes\ stars  in binaries. 
Companions of \hes\ stars are rapidly rotating. As is noted by \citet{2022MNRAS.516.3602E}, measurements of the radial velocity shifts of Be-stars are not reliable due to their high
rotation velocities and disk-driven spectral variability. Therefore, measuring the orbital
periods of binaries containing HeS stars is not easy. 
If taken at face value, in the sample of known Galactic Be-stars with identified sdO companions, the orbital periods do not exceed $\simeq$ 200\,days. 

The dependence of HeS star parameters on the effective temperature is presented in the left panel of Fig.~\ref{f:He_H_over}. 
In the  $\mathrm{nHe/nH} - {\rm T_{eff}}$ diagram, two groups 
of stars can be seen. The upper, more populous group is formed by stars that overflew critical lobes at $\porb \aplt 100$\,day. Initially wider systems are predominantly He-poor since they have  
heavier He cores and less H-exhausted surface layers when RLOF terminates.   The lower group includes stars that retained envelopes more massive than 0.3\,\ms; these stars form in binaries with the longest initial 
\porb. The stars in this group are the most massive among the HeS star population. 
Actually, this plot reflects the fact that more massive stars need to lose relatively less matter in order to detach from critical lobes, as was noticed already in early papers on case B evolution in massive stars
\citep[e.g., ][]{1973NInfo..27....3T}. Compared to the sdB/O stars, the range of nHe/nH of \hes\ stars is much more narrow; for the  former, its logarithm ranges from -4.0 to  3.0 \citep[][Fig.5]{2016PASP..128h2001H}. 

The $\log(\te) - \log(g)$ diagram is also structured. Evidently, the most compact and massive stars have the largest surface gravities. There is a branch of stars stretching from a high $g$, high \te\ region to a low $g$,  low \te\ region formed by relatively less massive \hes\ stars. This may be understood as an effect of 
RLOF at increasing periods, resulting in less compact \hes\ stars. The $\log({\rm g})$ values of \hes\ stars are within a broad range from 5.5 to 2.5, while for sdB/O stars they fall inside  the narrower (6.3 -- 5.0) interval 
\citep[][Fig.5]{2016PASP..128h2001H}. This reflects a broader range of masses and radii of \hes\ stars compared to subdwarfs. 

The right panel of Figure~\ref{f:He_H_over} shows the distribution of masses of H/He envelopes retained by \hes\ stars  at the lowest luminosity points along the evolutionary tracks in the synthesized population of \hes\ stars.
Most \hes\ stars have low-mass envelopes with $-0.2 \aplt \log(\rm{nHe/nH}) \aplt 0.2$, and hence the envelopes of these stars are He-dominated.
The profile of chemical abundances in the envelopes
of nascent \hes\ stars is defined by several factors: the stellar mass at the RLOF termination, the profile of abundances in the star, and the stellar wind acting as the star moves to the high \te\ region of the HRD. More than half of \hes\ stars descend from low-mass progenitors, which have relatively massive H/He envelopes after the cessation of RLOF. In this case, the increase in nHe/nH is due to the mass loss by stellar wind. 

The rest of the stars have a wider range of envelope masses, up to almost 1\,\ms\ in the most massive ones with a \te\ lower than 40\,000\,K. Though, in this group of stars also, the majority of objects have $\rm{M_{env}} \aplt 0.2$\,\ms\ (the darkest shades of the gray scale.) A more scattered group of stars with (0.2 -- 0.4)\,\ms\ envelopes extending toward high \te are
the remnants of more massive stars for which wind mass loss is less significant because of a very fast evolution. The X/Y ratio range of about 0.3 to 0.5 
is typical for the remnants of massive donors in close binaries.

The middle panel demonstrates an evident fact, that stars with low-mass envelopes are more compact, and hence have a higher $\log(g)$.

Finally, the HRD in the lower panel shows, again, a kind of compact sequence of descendants of low-mass (in the range under study) stars with the least massive envelopes and a scattered population of stars with different envelope masses depending on the initial masses of \hes\ star progenitors. Remarkably, the most massive envelopes have \hes\ stars positioned in the HRD in the region close to $\log(\te) \approx 4.4$, where the domain of \hes\ stars and WR stars 
overlaps with the main sequence (Figs.~\ref{f:HR_1}, \ref{f:HR_2}.)

\section{Discussion and conclusion} 
\label{sec:disc}
In the present paper we have addressed the issue of the observed scarcity of Galactic hot helium stars (HeS) with masses in the range (2--7)\,\ms, that is, between the most massive sdO subdwarfs and WR stars. We have performed a population synthesis of such stars, based on evolutionary computations for a grid of close binary systems leading to the formation of \hes\ stars  in this mass range, using the code MESA. 
We explored the full range
 of initial masses, orbital periods, and mass ratios of components of close binaries resulting in the formation of hot He stars in the above-mentioned mass range (Figs.~\ref{f:M_Porb_full_grid}, 
 \ref{f:fig15_4_2log}.) In this sense, our investigation is more comprehensive than the studies that aimed at the same stars \citep{2018A&A...615A..78G} or at the formation of progenitors of core-collapse SNe that pass through the same evolutionary stage and have masses in the same range as \hes\ star progenitors, for example \citet{2017ApJ...840...10Y} and \citet{2020ApJ...903...70S}.
We took into account rotation-induced mixing for the most tight binaries experiencing case A mass exchange. 
 
The results of our population synthesis suggest that there can be several reasons for the apparent scarcity of \hes\ binaries in the Galaxy.
In the first place, to form a \hes\ star, a stable mass transfer in a binary should occur, which is possible in restricted ranges of the initial orbital periods, $P_0$, and binary mass ratios, $q_0$, allowing the avoidance of a runaway mass transfer and common envelope formation. Our calculations enabled us to find the corresponding range of the initial binary systems' parameters (see Fig. 1.) A detailed tracing of the evolutionary paths leading to the \hes\ stars' location in the HRD (Figs.~\ref{f:HR_1}, \ref{f:HR_2}) suggests that most of the Galactic \hes\ stars descend from the initial binaries with ZAMS primary masses from $\simeq 5 M_\odot$ to $\simeq 24 M_\odot$. 

The remnants of stars that stably lose mass via RLOF may still retain massive H/He envelopes and never become ``hot'' (they have $\logte < 4.4$.) For the assumed Galactic SFR=$2 M_\odot$~yr$^{-1}$, our systematic exploration of the entire range of possible progenitors of He-stars with masses greater than 1\,\ms\ resulted in a Galactic number of \hes\  stars close to 
 20\,000 or about  3\,000,
if their mass exceeds $\sim 2$\,\ms\ -- the mass of the most heavy subdwarfs (Fig.~\ref{m1_distr}.)  

\citet{2018A&A...615A..78G} evaluated the fraction of Galactic early B- and O-stars hiding \hes\ stars companions as $\sim$3\%,
assuming that 1/3 of all massive 
stars in the Galaxy may produce \hes\ stars, if they overflow their Roche lobes 
before completion of stable core He burning. 
This number may be an overestimate. For 
stars more massive than 15\,\ms, the formation of \hes\ stars is possible 
in case B provided that they exhausted no more than 10-20 percent of the core He. The critical initial binary mass ratio, $q_0$, that still allows a stable mass exchange hardly exceeds $\sim 0.4$.
Additionally, for the avoidance of runaway mass loss and the formation of common envelopes with the further merger of companions, the progenitor binaries of \hes\ stars should not have initial orbital periods exceeding several hundred days.
 
We note that hot He-stars may have not only massive, but also intermediate-mass companions (see Fig.~\ref{f:fig15_4_2log}.) For example, out of a system with a minimum \hes\ star progenitor's mass of 7\,\ms\ and $q_0\geq0.6$, a binary with (1+4)\,\ms\ can be formed, and a 4 \ms\ star would appear as an early A-star.   

In our analysis, we have discarded low-mass helium remnants cooler than $\te \approx 20\,000\,K$, but we plan to discuss this problem in the future. 

Short-period binaries subject to case A mass exchange, with initial orbital periods, ${\rm P_0}$=(2 -- 5) days, may also give rise to \hes\ stars. Since IMF and distribution over ${\rm P_0}$ favor their formation, they contribute about 30 percent of all \hes\ stars. 
Formation of \hes\ stars in case A is possible since in this case  precursors of HeS stars are rapid rotators and rotation-induced effects reduce radius expansion of the main-sequence stars in very close binaries compared to more slowly rotating components in wider binaries \citep{2000ApJ...544.1016H}. The rotation effects allow some binary components to avoid the potential contact and possible coalescence on the main sequence.

We note that hot helium stars with masses (1 --10)\,\ms\ are supposed to experience short-period radial pulsations
accompanied by the formation of periodic shock waves \citep{2003AstL...29..522F}. More massive WR stars experience radial and non-radial pulsations as well \citep[e.g.,][]{2021MNRAS.502.5038N}. We
plan to investigate pulsations of the intermediate-mass evolved helium stars in a separate paper (Fadeyev et al., {\it in preparation}.) 

Selection effects reduce the number of potentially observable Galactic \hes\ stars. In the visual range, a hot He star emitting mostly in UV is extremely
difficult to discover in a binary, because its companion is a cooler and brighter B or Be-star (Fig.\ref{f:fig15_4_2log}.) This effect will reduce the number of observable \hes\ stars by at least a factor of two (Fig.~\ref{f:fig15_4_2log}.) More detailed spectral model calculations are required to improve these estimates using the atmospheric parameters and chemical composition of the synthesized \hes\ population (Fig.~\ref{f:He_H_over}.)

In our study, we have assumed that stellar winds of \hes\ stars obey the empirical \citet{2000A&A...360..227N} law. However, the issue of stellar winds from hot helium stars is not solved as of yet. \citet{2017A&A...607L...8V} suggested a model of radiation-driven mass loss wind for He stars that predicts mass loss rates by an order of magnitude lower than Nugis \& Lamers' law. \citet{2023arXiv230700074G} claim that some of the stripped stars discovered in the SMC obey Vink's law, while some of them possess even weaker winds. As the problem remains unsolved, we performed several test calculations comparing the influence of Nugis-Lamers' and Vink's winds upon the
masses of retained H/He envelopes and their chemical composition. Plots with results are presented in Appendixes A and B. As could be expected, Nugis-Lamers winds 
result in lower masses of envelopes at the He-burning part of the evolutionary tracks and lower He abundances. However, the difference in the envelope masses for considered models is only (5 -- 7)\%.

We conclude that, according to present binary evolution models, there are expected to be a few thousand HeS stars in massive binary systems in the Galaxy. Still, this population of hot He-rich stars remains hidden. Future work to better understand the selection effects and predict the observational signatures of HeS stars in binaries is needed to tailor observing campaigns aimed at discovering these elusive products of binary evolution.   

\begin{acknowledgements} 
The authors thank the anonymous referee for useful notes and suggestions.
The authors acknowledge fruitful discussions with Dr. N. Chugai.
L. Yungelson  acknowledges support via the German Academic Exchange Service DAAD  Research Stays for University Academics and Scientists Program.
A. Kuranov acknowledges support by the Russian Science Foundation grant 21-12-00141 (calculations of evolutionary models using the MESA code.)
This research has made use of the NASA’s Astrophysics Data System Bibliographic Services and of the SIMBAD database, operated at CDS, Strasbourg, France. 
\end{acknowledgements}

\bibliographystyle{aa}
\bibliography{./main}{}

\begin{thebibliography}{98}
\expandafter\ifx\csname natexlab\endcsname\relax\def\natexlab#1{#1}\fi

\bibitem[{{Barbaro} {et~al.}(1969){Barbaro}, {Giannone}, {Giannuzzi}, \&
  {Summa}}]{1969ASSL...13..217B}
{Barbaro}, G., {Giannone}, P., {Giannuzzi}, M.~A., \& {Summa}, C. 1969, in
  Astrophysics and Space Science Library, Vol.~13, Mass Loss from Stars, ed.
  M.~{Hack}, 217

\bibitem[{{Barr{\'\i}a} {et~al.}(2013){Barr{\'\i}a}, {Mennickent},
  {Schmidtobreick}, {Djura{\v{s}}evi{\'c}}, {Ko{\l}aczkowski}, {Michalska},
  {Vu{\v{c}}kovi{\'c}}, \& {Niemczura}}]{2013A&A...552A..63B}
{Barr{\'\i}a}, D., {Mennickent}, R.~E., {Schmidtobreick}, L., {et~al.} 2013,
  \aap, 552, A63

\bibitem[{{Bodensteiner} {et~al.}(2020){Bodensteiner}, {Shenar}, {Mahy},
  {Fabry}, {Marchant}, {Abdul-Masih}, {Banyard}, {Bowman}, {Dsilva}, {Frost},
  {Hawcroft}, {Reggiani}, \& {Sana}}]{2020A&A...641A..43B}
{Bodensteiner}, J., {Shenar}, T., {Mahy}, L., {et~al.} 2020, \aap, 641, A43

\bibitem[{{Chanlaridis} {et~al.}(2022){Chanlaridis}, {Antoniadis},
  {Aguilera-Dena}, {Gr{\"a}fener}, {Langer}, \&
  {Stergioulas}}]{2022A&A...668A.106C}
{Chanlaridis}, S., {Antoniadis}, J., {Aguilera-Dena}, D.~R., {et~al.} 2022,
  \aap, 668, A106

\bibitem[{{Chomiuk} \& {Povich}(2011)}]{2011AJ....142..197C}
{Chomiuk}, L. \& {Povich}, M.~S. 2011, \aj, 142, 197

\bibitem[{{Crowther}(2015)}]{2015wrs..conf...21C}
{Crowther}, P.~A. 2015, in Wolf-Rayet Stars, ed. W.-R. {Hamann}, A.~{Sander},
  \& H.~{Todt}, 21--26

\bibitem[{{de Jager} {et~al.}(1988){de Jager}, {Nieuwenhuijzen}, \& {van der
  Hucht}}]{1988A&AS...72..259D}
{de Jager}, C., {Nieuwenhuijzen}, H., \& {van der Hucht}, K.~A. 1988, \aaps,
  72, 259

\bibitem[{{de Kool}(1990)}]{1990ApJ...358..189D}
{de Kool}, M. 1990, \apj, 358, 189

\bibitem[{{De Loore} {et~al.}(1974){De Loore}, {de Gr{\`e}ve}, {van den
  Heuvel}, \& {de Cuyper}}]{1974MmSAI..45..893D}
{De Loore}, C., {de Gr{\`e}ve}, J.~P., {van den Heuvel}, E.~P.~J., \& {de
  Cuyper}, J.~P. 1974, \memsai, 45, 893

\bibitem[{{de Mink} {et~al.}(2011){de Mink}, {Langer}, \&
  {Izzard}}]{2011IAUS..272..531D}
{de Mink}, S.~E., {Langer}, N., \& {Izzard}, R.~G. 2011, in Active OB Stars:
  Structure, Evolution, Mass Loss, and Critical Limits, ed. C.~{Neiner},
  G.~{Wade}, G.~{Meynet}, \& G.~{Peters}, Vol. 272, 531--532

\bibitem[{{Dessart} {et~al.}(2020){Dessart}, {Yoon}, {Aguilera-Dena}, \&
  {Langer}}]{2020A&A...642A.106D}
{Dessart}, L., {Yoon}, S.-C., {Aguilera-Dena}, D.~R., \& {Langer}, N. 2020,
  \aap, 642, A106

\bibitem[{{Dionne} \& {Robert}(2006)}]{Dionne2006}
{Dionne}, D. \& {Robert}, C. 2006, \apj, 641, 252

\bibitem[{{Doughty} \& {Finlator}(2021)}]{Doughty2021}
{Doughty}, C. \& {Finlator}, K. 2021, \mnras, 505, 2207

\bibitem[{{Drout} {et~al.}(2023){Drout}, {G{\"o}tberg}, {Ludwig}, {Groh}, {de
  Mink}, {O'Grady}, \& {Smith}}]{2023arXiv230700061D}
{Drout}, M.~R., {G{\"o}tberg}, Y., {Ludwig}, B.~A., {et~al.} 2023, arXiv
  e-prints, arXiv:2307.00061

\bibitem[{{Ekstr{\"o}m} {et~al.}(2012){Ekstr{\"o}m}, {Georgy}, {Eggenberger},
  {Meynet}, {Mowlavi}, {Wyttenbach}, {Granada}, {Decressin}, {Hirschi},
  {Frischknecht}, {Charbonnel}, \& {Maeder}}]{2012A&A...537A.146E}
{Ekstr{\"o}m}, S., {Georgy}, C., {Eggenberger}, P., {et~al.} 2012, \aap, 537,
  A146

\bibitem[{{El-Badry} \& {Burdge}(2022)}]{2022MNRAS.511L..24E}
{El-Badry}, K. \& {Burdge}, K.~B. 2022, \mnras, 511, 24

\bibitem[{{El-Badry} {et~al.}(2022){El-Badry}, {Conroy}, {Quataert}, {Rix},
  {Labadie-Bartz}, {Jayasinghe}, {Thompson}, {Cargile}, {Stassun}, \&
  {Ilyin}}]{2022MNRAS.516.3602E}
{El-Badry}, K., {Conroy}, C., {Quataert}, E., {et~al.} 2022, \mnras, 516, 3602

\bibitem[{{Fadeyev} \& {Novikova}(2003)}]{2003AstL...29..522F}
{Fadeyev}, Y.~A. \& {Novikova}, M.~F. 2003, Astronomy Letters, 29, 522

\bibitem[{{Gagnier} \& {Pejcha}(2023)}]{2023A&A...674A.121G}
{Gagnier}, D. \& {Pejcha}, O. 2023, \aap, 674, A121

\bibitem[{{Geier}(2020)}]{2020A&A...635A.193G}
{Geier}, S. 2020, \aap, 635, A193

\bibitem[{{Giannone} \& {Giannuzzi}(1972)}]{1972A&A....19..298G}
{Giannone}, P. \& {Giannuzzi}, M.~A. 1972, \aap, 19, 298

\bibitem[{{Giannone} {et~al.}(1970){Giannone}, {Refsdal}, \&
  {Weigert}}]{1970A&A.....4..428G}
{Giannone}, P., {Refsdal}, S., \& {Weigert}, A. 1970, \aap, 4, 428

\bibitem[{{Gies} {et~al.}(2023){Gies}, {Wang}, \&
  {Klement}}]{2023ApJ...942L...6G}
{Gies}, D.~R., {Wang}, L., \& {Klement}, R. 2023, \apjl, 942, L6

\bibitem[{{G{\"o}tberg} {et~al.}(2017){G{\"o}tberg}, {de Mink}, \&
  {Groh}}]{2017A&A...608A..11G}
{G{\"o}tberg}, Y., {de Mink}, S.~E., \& {Groh}, J.~H. 2017, \aap, 608, A11

\bibitem[{{G\"{o}tberg} {et~al.}(2018){G\"{o}tberg}, {de Mink}, {Groh},
  {Kupfer}, {Crowther}, {Zapartas}, \& {Renzo}}]{2018A&A...615A..78G}
{G\"{o}tberg}, Y., {de Mink}, S.~E., {Groh}, J.~H., {et~al.} 2018, \aap, 615,
  A78

\bibitem[{{G\"{o}tberg} {et~al.}(2023){G\"{o}tberg}, {Drout}, {Ji}, {Groh},
  {Ludwig}, {Crowther}, {Smith}, {de Koter}, \& {de
  Mink}}]{2023arXiv230700074G}
{G\"{o}tberg}, Y., {Drout}, M.~R., {Ji}, A.~P., {et~al.} 2023, arXiv e-prints,
  arXiv:2307.00074

\bibitem[{{Habets}(1986)}]{1986A&A...167...61H}
{Habets}, G.~M.~H.~J. 1986, \aap, 167, 61

\bibitem[{{Hainich} {et~al.}(2019){Hainich}, {Ramachandran}, {Shenar},
  {Sander}, {Todt}, {Gruner}, {Oskinova}, \& {Hamann}}]{2019A&A...621A..85H}
{Hainich}, R., {Ramachandran}, V., {Shenar}, T., {et~al.} 2019, \aap, 621, A85

\bibitem[{{Hamann} {et~al.}(2019){Hamann}, {Gr{\"a}fener}, {Liermann},
  {Hainich}, {Sander}, {Shenar}, {Ramachandran}, {Todt}, \&
  {Oskinova}}]{2019A&A...625A..57H}
{Hamann}, W.~R., {Gr{\"a}fener}, G., {Liermann}, A., {et~al.} 2019, \aap, 625,
  A57

\bibitem[{{Han} {et~al.}(2003){Han}, {Podsiadlowski}, {Maxted}, \&
  {Marsh}}]{2003MNRAS.341..669H}
{Han}, Z., {Podsiadlowski}, P., {Maxted}, P.~F.~L., \& {Marsh}, T.~R. 2003,
  \mnras, 341, 669

\bibitem[{{Han} {et~al.}(2002){Han}, {Podsiadlowski}, {Maxted}, {Marsh}, \&
  {Ivanova}}]{2002MNRAS.336..449H}
{Han}, Z., {Podsiadlowski}, P., {Maxted}, P.~F.~L., {Marsh}, T.~R., \&
  {Ivanova}, N. 2002, \mnras, 336, 449

\bibitem[{{Harmanec}(1970)}]{1970Ap&SS...6..497H}
{Harmanec}, P. 1970, \apss, 6, 497

\bibitem[{{Heber}(2016)}]{2016PASP..128h2001H}
{Heber}, U. 2016, \pasp, 128, 082001

\bibitem[{{Heger} \& {Langer}(2000)}]{2000ApJ...544.1016H}
{Heger}, A. \& {Langer}, N. 2000, \apj, 544, 1016

\bibitem[{{Heger} {et~al.}(2000){Heger}, {Langer}, \&
  {Woosley}}]{2000ApJ...528..368H}
{Heger}, A., {Langer}, N., \& {Woosley}, S.~E. 2000, \apj, 528, 368

\bibitem[{{Hennicker} {et~al.}(2022){Hennicker}, {Kee}, {Shenar},
  {Bodensteiner}, {Abdul-Masih}, {El Mellah}, {Sana}, \&
  {Sundqvist}}]{2022A&A...660A..17H}
{Hennicker}, L., {Kee}, N.~D., {Shenar}, T., {et~al.} 2022, \aap, 660, A17

\bibitem[{{Howarth} \& {Heber}(1990)}]{1990PASP..102..912H}
{Howarth}, I.~D. \& {Heber}, U. 1990, \pasp, 102, 912

\bibitem[{{Iben} \& {Tutukov}(1985)}]{1985ApJS...58..661I}
{Iben}, I., J. \& {Tutukov}, A.~V. 1985, \apjs, 58, 661

\bibitem[{{Iben} \& {Tutukov}(1987)}]{1987ApJ...313..727I}
{Iben}, Icko, J. \& {Tutukov}, A.~V. 1987, \apj, 313, 727

\bibitem[{{Irrgang} {et~al.}(2022){Irrgang}, {Przybilla}, \&
  {Meynet}}]{2022NatAs...6.1414I}
{Irrgang}, A., {Przybilla}, N., \& {Meynet}, G. 2022, Nature Astronomy, 6, 1414

\bibitem[{{Ivanova} {et~al.}(2013){Ivanova}, {Justham}, {Chen}, {De Marco},
  {Fryer}, {Gaburov}, {Ge}, {Glebbeek}, {Han}, {Li}, {Lu}, {Marsh},
  {Podsiadlowski}, {Potter}, {Soker}, {Taam}, {Tauris}, {van den Heuvel}, \&
  {Webbink}}]{2013A&ARv..21...59I}
{Ivanova}, N., {Justham}, S., {Chen}, X., {et~al.} 2013, \aapr, 21, 59

\bibitem[{{Kanarek}(2017)}]{2017PhDT.......418K}
{Kanarek}, G.~C. 2017, PhD thesis, Columbia University, New York

\bibitem[{{Kim} {et~al.}(2015){Kim}, {Yoon}, \& {Koo}}]{2015ApJ...809..131K}
{Kim}, H.-J., {Yoon}, S.-C., \& {Koo}, B.-C. 2015, \apj, 809, 131

\bibitem[{{Kippenhahn}(1969)}]{1969A&A.....3...83K}
{Kippenhahn}, R. 1969, \aap, 3, 83

\bibitem[{{Kippenhahn} {et~al.}(1967){Kippenhahn}, {Kohl}, \&
  {Weigert}}]{1967ZA.....66...58K}
{Kippenhahn}, R., {Kohl}, K., \& {Weigert}, A. 1967, \zap, 66, 58

\bibitem[{{Kippenhahn} \& {Meyer-Hofmeister}(1977)}]{1977A&A....54..539K}
{Kippenhahn}, R. \& {Meyer-Hofmeister}, E. 1977, \aap, 54, 539

\bibitem[{{Kippenhahn} \& {Weigert}(1967)}]{1967ZA.....65..251K}
{Kippenhahn}, R. \& {Weigert}, A. 1967, \zap, 65, 251

\bibitem[{{Klement} {et~al.}(2022){Klement}, {Baade}, {Rivinius}, {Gies},
  {Wang}, {Labadie-Bartz}, {Ticiani dos Santos}, {Monnier}, {Carciofi},
  {M{\'e}rand}, {Anugu}, {Schaefer}, {Le Bouquin}, {Davies}, {Ennis},
  {Gardner}, {Kraus}, {Setterholm}, \& {Labdon}}]{2022ApJ...940...86K}
{Klement}, R., {Baade}, D., {Rivinius}, T., {et~al.} 2022, \apj, 940, 86

\bibitem[{{Klencki} {et~al.}(2022){Klencki}, {Istrate}, {Nelemans}, \&
  {Pols}}]{2022A&A...662A..56K}
{Klencki}, J., {Istrate}, A., {Nelemans}, G., \& {Pols}, O. 2022, \aap, 662,
  A56

\bibitem[{{Kriz} \& {Harmanec}(1975)}]{1975BAICz..26...65K}
{Kriz}, S. \& {Harmanec}, P. 1975, Bulletin of the Astronomical Institutes of
  Czechoslovakia, 26, 65

\bibitem[{{Langer} {et~al.}(2003){Langer}, {Wellstein}, \&
  {Petrovic}}]{2003IAUS..212..275L}
{Langer}, N., {Wellstein}, S., \& {Petrovic}, J. 2003, in A Massive Star
  Odyssey: From Main Sequence to Supernova, ed. K.~{van der Hucht},
  A.~{Herrero}, \& C.~{Esteban}, Vol. 212, 275

\bibitem[{{Lauterborn}(1970)}]{1970A&A.....7..150L}
{Lauterborn}, D. 1970, \aap, 7, 150

\bibitem[{{Licquia} \& {Newman}(2015)}]{2015ApJ...806...96L}
{Licquia}, T.~C. \& {Newman}, J.~A. 2015, \apj, 806, 96

\bibitem[{{Liu} {et~al.}(2019){Liu}, {Zhang}, {Howard}, {Bai}, {Lu}, {Soria},
  {Justham}, {Li}, {Zheng}, {Wang}, {Belczynski}, {Casares}, {Zhang}, {Yuan},
  {Dong}, {Lei}, {Isaacson}, {Wang}, {Bai}, {Shao}, {Gao}, {Wang}, {Niu},
  {Cui}, {Zheng}, {Mu}, {Zhang}, {Wang}, {Heger}, {Qi}, {Liao}, {Lattanzi},
  {Gu}, {Wang}, {Wu}, {Shao}, {Shen}, {Wang}, {Bregman}, {Di Stefano}, {Liu},
  {Han}, {Zhang}, {Wang}, {Ren}, {Zhang}, {Zhang}, {Wang}, {Cabrera-Lavers},
  {Corradi}, {Rebolo}, {Zhao}, {Zhao}, {Chu}, \& {Cui}}]{2019Natur.575..618L}
{Liu}, J., {Zhang}, H., {Howard}, A.~W., {et~al.} 2019, \nat, 575, 618

\bibitem[{{Menon} {et~al.}(2021){Menon}, {Langer}, {de Mink}, {Justham}, {Sen},
  {Sz{\'e}csi}, {de Koter}, {Abdul-Masih}, {Sana}, {Mahy}, \&
  {Marchant}}]{2021MNRAS.507.5013M}
{Menon}, A., {Langer}, N., {de Mink}, S.~E., {et~al.} 2021, \mnras, 507, 5013

\bibitem[{{Naz{\'e}} {et~al.}(2021){Naz{\'e}}, {Rauw}, \&
  {Gosset}}]{2021MNRAS.502.5038N}
{Naz{\'e}}, Y., {Rauw}, G., \& {Gosset}, E. 2021, \mnras, 502, 5038

\bibitem[{{Nomoto} {et~al.}(1994){Nomoto}, {Yamaoka}, {Pols}, {van den Heuvel},
  {Iwamoto}, {Kumagai}, \& {Shigeyama}}]{1994Natur.371..227N}
{Nomoto}, K., {Yamaoka}, H., {Pols}, O.~R., {et~al.} 1994, \nat, 371, 227

\bibitem[{{Nugis} \& {Lamers}(2000)}]{2000A&A...360..227N}
{Nugis}, T. \& {Lamers}, H.~J.~G.~L.~M. 2000, \aap, 360, 227

\bibitem[{{Ohlmann} {et~al.}(2016){Ohlmann}, {R{\"o}pke}, {Pakmor}, \&
  {Springel}}]{2016ApJ...816L...9O}
{Ohlmann}, S.~T., {R{\"o}pke}, F.~K., {Pakmor}, R., \& {Springel}, V. 2016,
  \apjl, 816, L9

\bibitem[{{{\"O}pik}(1924)}]{1924PTarO..25f...1O}
{{\"O}pik}, E. 1924, Publications of the Tartu Astrofizica Observatory, 25, 1

\bibitem[{{Packet}(1981)}]{1981A&A...102...17P}
{Packet}, W. 1981, \aap, 102, 17

\bibitem[{{Paczy{\'n}ski}(1967)}]{1967AcA....17..355P}
{Paczy{\'n}ski}, B. 1967, \actaa, 17, 355

\bibitem[{{Paxton} {et~al.}(2011){Paxton}, {Bildsten}, {Dotter}, {Herwig},
  {Lesaffre}, \& {Timmes}}]{2011ApJS..192....3P}
{Paxton}, B., {Bildsten}, L., {Dotter}, A., {et~al.} 2011, \apjs, 192, 3

\bibitem[{{Paxton} {et~al.}(2013){Paxton}, {Cantiello}, {Arras}, {Bildsten},
  {Brown}, {Dotter}, {Mankovich}, {Montgomery}, {Stello}, {Timmes}, \&
  {Townsend}}]{2013ApJS..208....4P}
{Paxton}, B., {Cantiello}, M., {Arras}, P., {et~al.} 2013, \apjs, 208, 4

\bibitem[{{Paxton} {et~al.}(2015){Paxton}, {Marchant}, {Schwab}, {Bauer},
  {Bildsten}, {Cantiello}, {Dessart}, {Farmer}, {Hu}, {Langer}, {Townsend},
  {Townsley}, \& {Timmes}}]{2015ApJS..220...15P}
{Paxton}, B., {Marchant}, P., {Schwab}, J., {et~al.} 2015, \apjs, 220, 15

\bibitem[{{Paxton} {et~al.}(2018){Paxton}, {Schwab}, {Bauer}, {Bildsten},
  {Blinnikov}, {Duffell}, {Farmer}, {Goldberg}, {Marchant}, {Sorokina},
  {Thoul}, {Townsend}, \& {Timmes}}]{2018ApJS..234...34P}
{Paxton}, B., {Schwab}, J., {Bauer}, E.~B., {et~al.} 2018, \apjs, 234, 34

\bibitem[{{Paxton} {et~al.}(2019){Paxton}, {Smolec}, {Schwab}, {Gautschy},
  {Bildsten}, {Cantiello}, {Dotter}, {Farmer}, {Goldberg}, {Jermyn}, {Kanbur},
  {Marchant}, {Thoul}, {Townsend}, {Wolf}, {Zhang}, \&
  {Timmes}}]{2019ApJS..243...10P}
{Paxton}, B., {Smolec}, R., {Schwab}, J., {et~al.} 2019, \apjs, 243, 10

\bibitem[{{Pecaut} \& {Mamajek}(2013)}]{2013ApJS..208....9P}
{Pecaut}, M.~J. \& {Mamajek}, E.~E. 2013, \apjs, 208, 9

\bibitem[{{Petrovic} {et~al.}(2005){Petrovic}, {Langer}, \& {van der
  Hucht}}]{2005A&A...435.1013P}
{Petrovic}, J., {Langer}, N., \& {van der Hucht}, K.~A. 2005, \aap, 435, 1013

\bibitem[{{Podsiadlowski}(1996)}]{1996ASPC...96..419P}
{Podsiadlowski}, P. 1996, in Astronomical Society of the Pacific Conference
  Series, Vol.~96, Hydrogen Deficient Stars, ed. C.~S. {Jeffery} \& U.~{Heber},
  419

\bibitem[{{Pols} {et~al.}(1991){Pols}, {Cote}, {Waters}, \&
  {Heise}}]{1991A&A...241..419P}
{Pols}, O.~R., {Cote}, J., {Waters}, L.~B.~F.~M., \& {Heise}, J. 1991, \aap,
  241, 419

\bibitem[{{Popova} {et~al.}(1982){Popova}, {Tutukov}, \&
  {Yungelson}}]{1982Ap&SS..88...55P}
{Popova}, E.~I., {Tutukov}, A.~V., \& {Yungelson}, L.~R. 1982, \apss, 88, 55

\bibitem[{{Ramachandran} {et~al.}(2023){Ramachandran}, {Klencki}, {Sander},
  {Pauli}, {Shenar}, {Oskinova}, \& {Hamann}}]{ramachandran2023partially}
{Ramachandran}, V., {Klencki}, J., {Sander}, A.~A.~C., {et~al.} 2023, \aap,
  674, L12

\bibitem[{{Refsdal} \& {Weigert}(1969)}]{1969ASSL...13..253R}
{Refsdal}, S. \& {Weigert}, A. 1969, in Astrophysics and Space Science Library,
  Vol.~13, Mass Loss from Stars, ed. M.~{Hack}, 253

\bibitem[{{Rosales Guzm{\'a}n} {et~al.}(2018){Rosales Guzm{\'a}n},
  {Mennickent}, {Djura{\v{s}}evi{\'c}}, {Araya}, \&
  {Cur{\'e}}}]{2018MNRAS.476.3039R}
{Rosales Guzm{\'a}n}, J.~A., {Mennickent}, R.~E., {Djura{\v{s}}evi{\'c}}, G.,
  {Araya}, I., \& {Cur{\'e}}, M. 2018, \mnras, 476, 3039

\bibitem[{{Sander} {et~al.}(2019){Sander}, {Hamann}, {Todt}, {Hainich},
  {Shenar}, {Ramachandran}, \& {Oskinova}}]{Sander2019}
{Sander}, A.~A.~C., {Hamann}, W.~R., {Todt}, H., {et~al.} 2019, \aap, 621, A92

\bibitem[{{Schaffenroth} {et~al.}(2022){Schaffenroth}, {Pelisoli}, {Barlow},
  {Geier}, \& {Kupfer}}]{2022A&A...666A.182S}
{Schaffenroth}, V., {Pelisoli}, I., {Barlow}, B.~N., {Geier}, S., \& {Kupfer},
  T. 2022, \aap, 666, A182

\bibitem[{{Sen} {et~al.}(2022){Sen}, {Langer}, {Marchant}, {Menon}, {de Mink},
  {Schootemeijer}, {Sch{\"u}rmann}, {Mahy}, {Hastings}, {Nathaniel}, {Sana},
  {Wang}, \& {Xu}}]{2022A&A...659A..98S}
{Sen}, K., {Langer}, N., {Marchant}, P., {et~al.} 2022, \aap, 659, A98

\bibitem[{{Shenar} {et~al.}(2020{\natexlab{a}}){Shenar}, {Bodensteiner},
  {Abdul-Masih}, {Fabry}, {Mahy}, {Marchant}, {Banyard}, {Bowman}, {Dsilva},
  {Hawcroft}, {Reggiani}, \& {Sana}}]{2020A&A...639L...6S}
{Shenar}, T., {Bodensteiner}, J., {Abdul-Masih}, M., {et~al.}
  2020{\natexlab{a}}, \aap, 639, L6

\bibitem[{{Shenar} {et~al.}(2020{\natexlab{b}}){Shenar}, {Gilkis}, {Vink},
  {Sana}, \& {Sander}}]{2020A&A...634A..79S}
{Shenar}, T., {Gilkis}, A., {Vink}, J.~S., {Sana}, H., \& {Sander}, A.~A.~C.
  2020{\natexlab{b}}, \aap, 634, A79

\bibitem[{{Shenar} {et~al.}(2023){Shenar}, {Wade}, {Marchant}, {Bagnulo},
  {Bodensteiner}, {Bowman}, {Gilkis}, {Langer}, {Nicolas-Chen{\'e}},
  {Oskinova}, {Van Reeth}, {Sana}, {St-Louis}, {de Oliveira}, {Todt}, \&
  {Toonen}}]{2023Sci...381..761S}
{Shenar}, T., {Wade}, G.~A., {Marchant}, P., {et~al.} 2023, Science, 381, 761

\bibitem[{{Sravan} {et~al.}(2020){Sravan}, {Marchant}, {Kalogera},
  {Milisavljevic}, \& {Margutti}}]{2020ApJ...903...70S}
{Sravan}, N., {Marchant}, P., {Kalogera}, V., {Milisavljevic}, D., \&
  {Margutti}, R. 2020, \apj, 903, 70

\bibitem[{{Tsujimoto} {et~al.}(2023){Tsujimoto}, {Hayashi}, {Morihana}, \&
  {Moritani}}]{2023PASJ...75..177T}
{Tsujimoto}, M., {Hayashi}, T., {Morihana}, K., \& {Moritani}, Y. 2023, \pasj,
  75, 177

\bibitem[{{Tutukov} \& {Iungelson}(1987)}]{1987fbs..conf..435T}
{Tutukov}, A. \& {Iungelson}, L. 1987, in IAU Colloq. 95: Second Conference on
  Faint Blue Stars, ed. A.~G.~D. {Philip}, D.~S. {Hayes}, \& J.~W. {Liebert},
  435--444

\bibitem[{{Tutukov} {et~al.}(1973){Tutukov}, {Yungelson}, \&
  {Klayman}}]{1973NInfo..27....3T}
{Tutukov}, A., {Yungelson}, L., \& {Klayman}, A. 1973, Nauchnye Informatsii,
  27, 3

\bibitem[{{Tutukov} \& {Yungelson}(1990)}]{1990SvA....34...57T}
{Tutukov}, A.~V. \& {Yungelson}, L.~R. 1990, \sovast, 34, 57

\bibitem[{{Uomoto}(1986)}]{1986ApJ...310L..35U}
{Uomoto}, A. 1986, \apjl, 310, L35

\bibitem[{{Villase{\~n}or} {et~al.}(2023){Villase{\~n}or}, {Lennon}, {Picco},
  {Shenar}, {Marchant}, {Langer}, {Dufton}, {Nardini}, {Evans}, {Bodensteiner},
  {de Mink}, {G{\"o}tberg}, {Soszy{\'n}ski}, {Taylor}, \&
  {Sana}}]{2023MNRAS.525.5121V}
{Villase{\~n}or}, J.~I., {Lennon}, D.~J., {Picco}, A., {et~al.} 2023, \mnras,
  525, 5121

\bibitem[{{Vink}(2017)}]{2017A&A...607L...8V}
{Vink}, J.~S. 2017, \aap, 607, L8

\bibitem[{{Vink} {et~al.}(2001){Vink}, {de Koter}, \&
  {Lamers}}]{2001A&A...369..574V}
{Vink}, J.~S., {de Koter}, A., \& {Lamers}, H.~J.~G.~L.~M. 2001, \aap, 369, 574

\bibitem[{{Waldman} {et~al.}(2008){Waldman}, {Yungelson}, \&
  {Barkat}}]{2008ASPC..391..359W}
{Waldman}, R., {Yungelson}, L.~R., \& {Barkat}, Z. 2008, in Astronomical
  Society of the Pacific Conference Series, Vol. 391, Hydrogen-Deficient Stars,
  ed. A.~{Werner} \& T.~{Rauch}, 359

\bibitem[{{Wang} {et~al.}(2023){Wang}, {Gies}, {Peters}, \&
  {Han}}]{2023AJ....165..203W}
{Wang}, L., {Gies}, D.~R., {Peters}, G.~J., \& {Han}, Z. 2023, \aj, 165, 203

\bibitem[{{Webbink}(1984)}]{1984ApJ...277..355W}
{Webbink}, R.~F. 1984, \apj, 277, 355

\bibitem[{{Yoon} {et~al.}(2017){Yoon}, {Dessart}, \&
  {Clocchiatti}}]{2017ApJ...840...10Y}
{Yoon}, S.-C., {Dessart}, L., \& {Clocchiatti}, A. 2017, \apj, 840, 10

\bibitem[{{Yoon} {et~al.}(2010){Yoon}, {Woosley}, \&
  {Langer}}]{2010ApJ...725..940Y}
{Yoon}, S.~C., {Woosley}, S.~E., \& {Langer}, N. 2010, \apj, 725, 940

\bibitem[{{Yungelson} \& {Tutukov}(2005)}]{2005ARep...49..871Y}
{Yungelson}, L.~R. \& {Tutukov}, A.~V. 2005, Astronomy Reports, 49, 871

\bibitem[{{Zak} {et~al.}(2023){Zak}, {Jones}, {Boffin}, {Beck}, {Klencki},
  {Bodensteiner}, {Shenar}, {Van Winckel}, {Skarka}, {Arellano-C{\'o}rdova},
  {Viuho}, {Sowicka}, {Guenther}, \& {Hatzes}}]{2023MNRAS.524.5749Z}
{Zak}, J., {Jones}, D., {Boffin}, H.~M.~J., {et~al.} 2023, \mnras, 524, 5749

\bibitem[{{Zi{\'o}{\l}kowski}(1970)}]{1970AcA....20..213Z}
{Zi{\'o}{\l}kowski}, J. 1970, \actaa, 20, 213

\end{thebibliography}
%%\newpage

\begin{appendix}

\section{Dependence of evolutionary tracks and masses of envelopes of hot helium stars on accepted stellar wind law}
\begin{figure}[h]
\includegraphics[width=8.4cm]{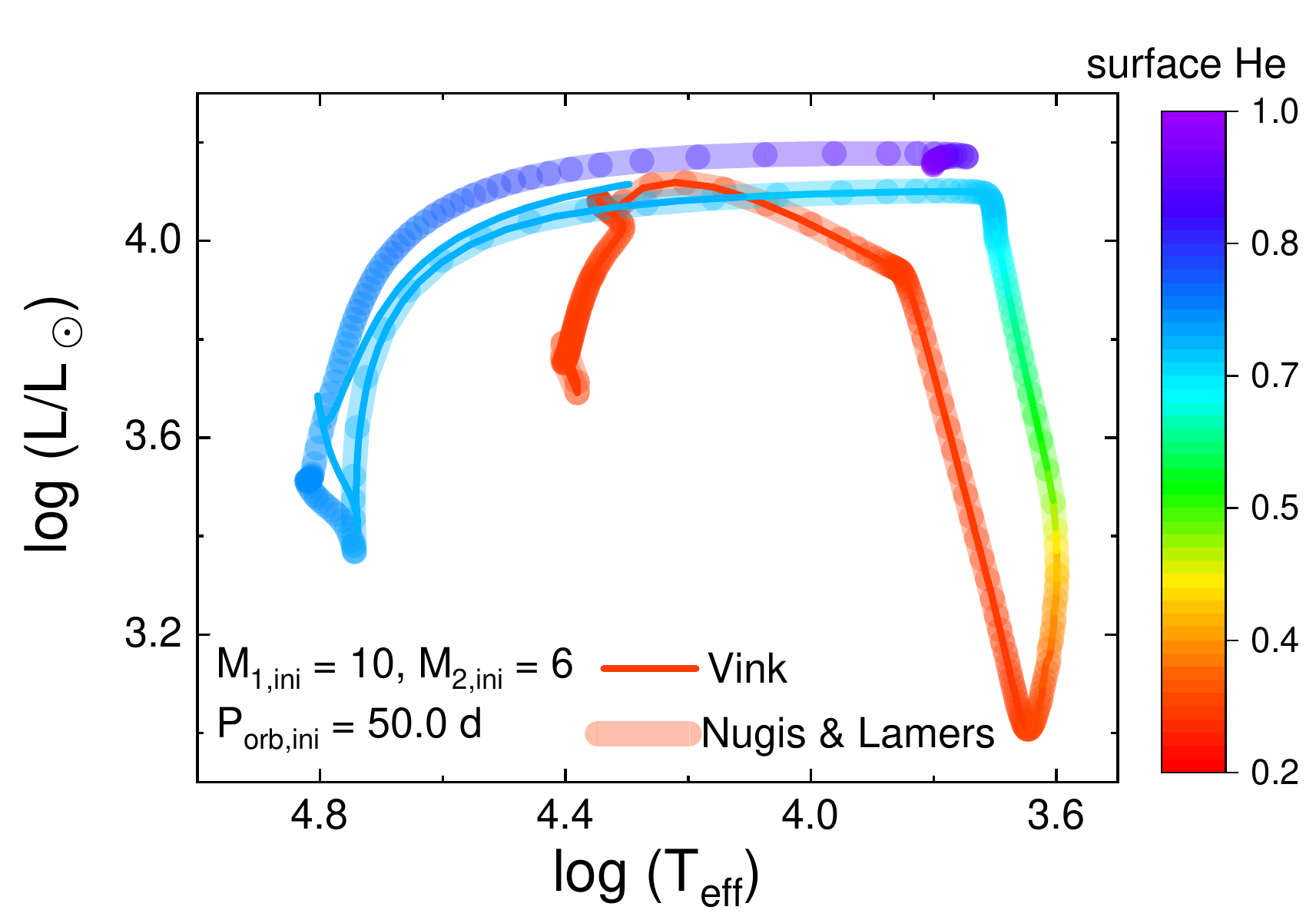}
\hskip -0.089cm
\includegraphics[width=8.4cm]{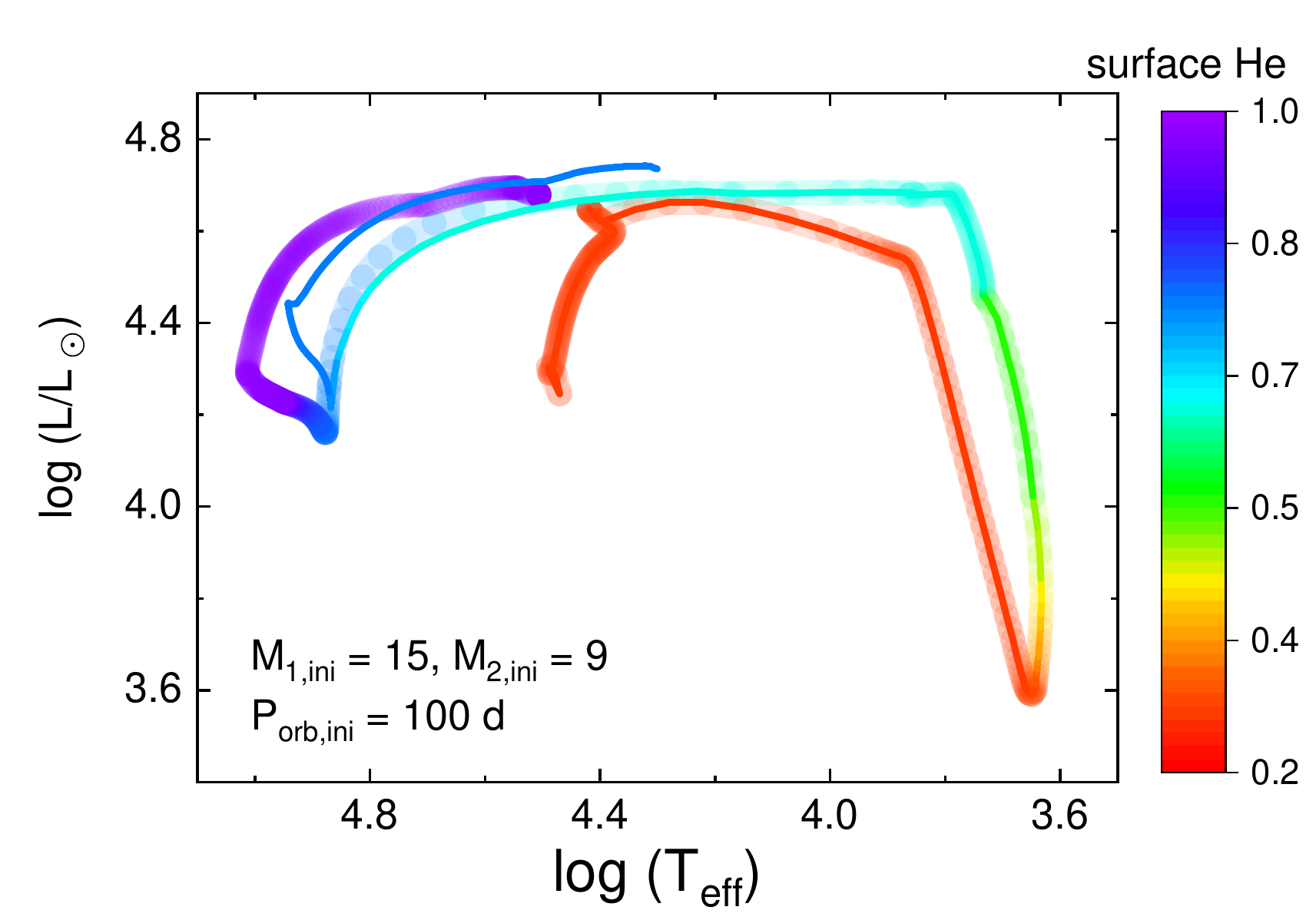}
%\resizebox{\hsize}{!}{\includegraphics{app_fig_v1.pdf}}
%\resizebox{\hsize}{!}{\includegraphics{app_fig_v2.pdf}}
\caption{HRD for the primary component of a 10+6 \ms\ binary with $P_0=50$ days 
({\it upper panel}) and a 15+9 \ms\ binary with ${\rm P_0}=100$ days ({\it lower panel}) forming \hes\ stars for wind mass-loss laws of \hes\ stars according to \citet{2000A&A...360..227N} and \citet{2017A&A...607L...8V} (the thin and thick lines, respectively.) The color scale to the right codes the surface He abundance Y.}
\label{f:a1}
\end{figure}

\newpage
\section{Dependence of the surface chemical composition of a hot helium star on the stellar wind law}

\begin{figure}[h]
%\centering
%\resizebox{\hsize}{!}{\includegraphics{app_fig_v3.pdf}}
%\resizebox{\hsize}{!}{\includegraphics{app_fig_v4.pdf}}
\includegraphics[width=8.4cm]{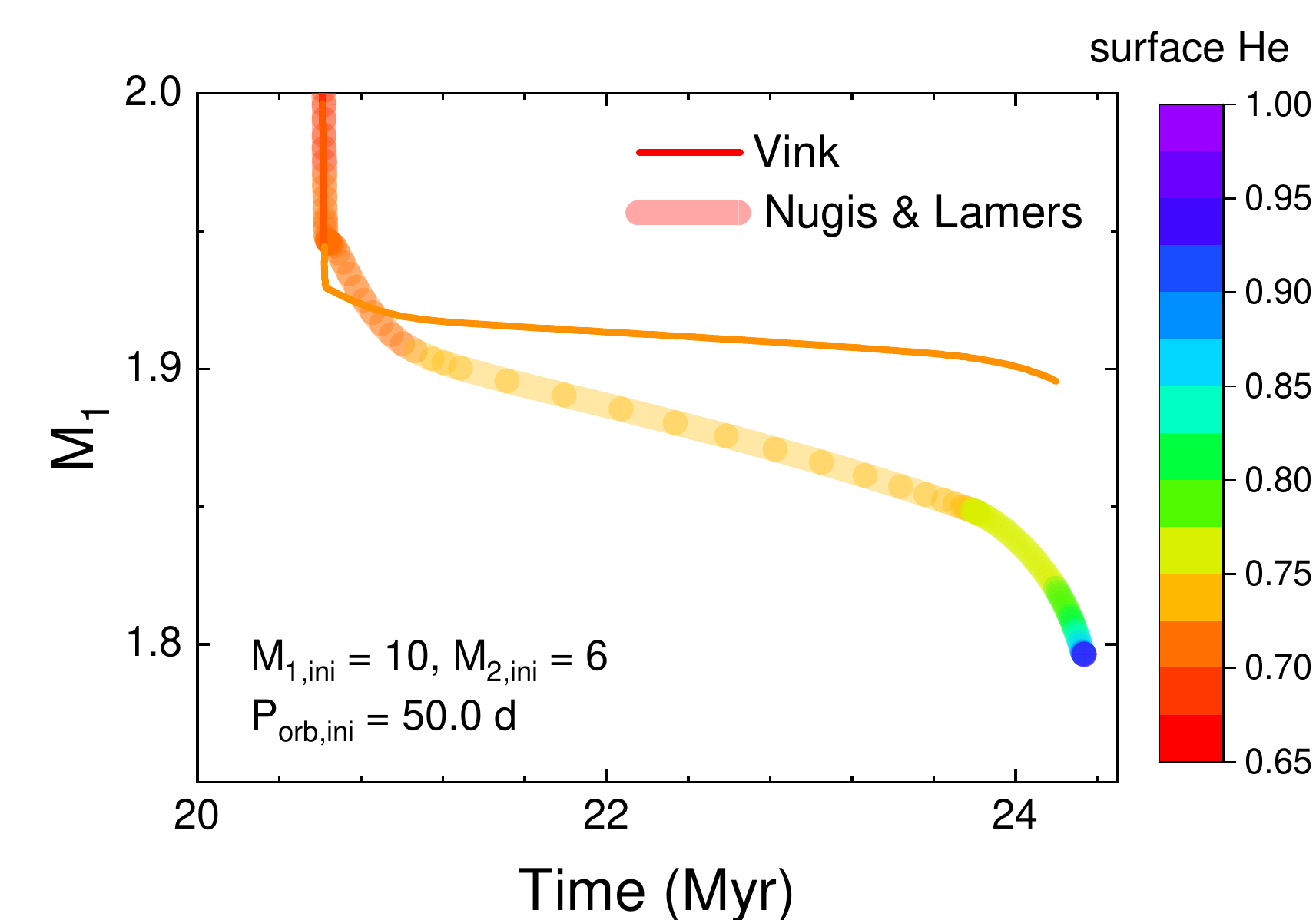}
\hskip -0.089cm
\includegraphics[width=8.4cm]{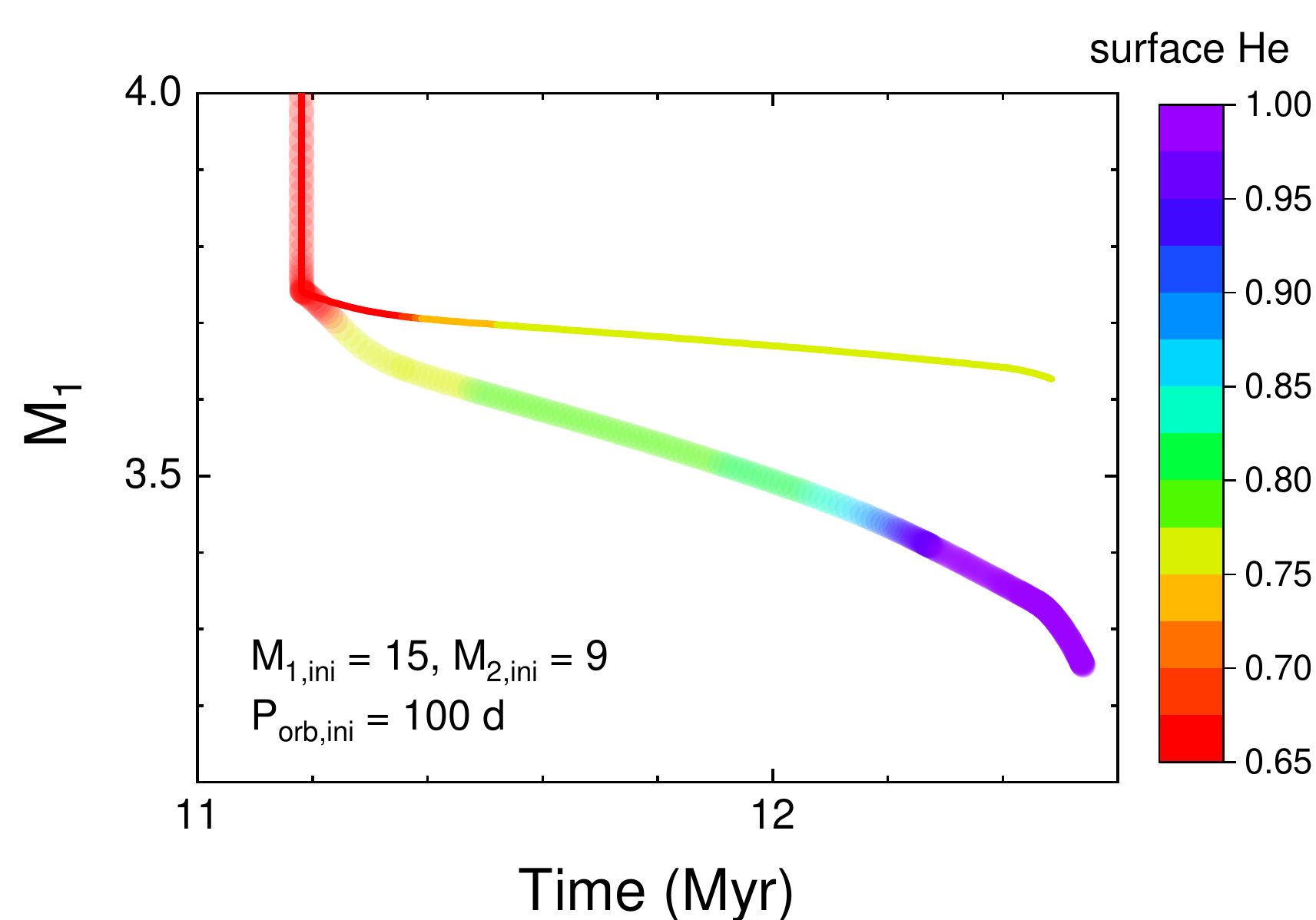}
\caption{Time dependence of the \hes\ mass formed from binaries shown in Fig. \ref{f:a1} for different \hes\ star stellar wind mass-loss laws. The color scale to the right codes the surface He abundance Y.}
\label{f:a2}
\end{figure}
\end{appendix}
\end{document}